\documentclass[aps,eqsecnum,twocolumn,showpacs,preprintnumbers,nofootinbib,prd,superscriptaddress,groupedaddress,10pt]{revtex4-2}

\makeatletter
\def\l@subsubsection#1#2{}
\def\l@subsubsubsection#1#2{}
\makeatother

\usepackage{empheq}
\usepackage{graphicx,epsfig,epsf,fixmath}
\usepackage[dvipsnames]{xcolor}
\usepackage{epstopdf}
\usepackage{mathtools}
\usepackage{amssymb}
\usepackage{mathrsfs}
\usepackage{relsize}
\usepackage{graphics}
\usepackage{placeins} 
\usepackage{booktabs} 
\usepackage{siunitx} 
\DeclareSIUnit \parsec {pc} 
\usepackage{aas_macros}
\usepackage{bm}
\usepackage{dcolumn}
\usepackage{latexsym}
\usepackage{rotating}
\usepackage{longtable} 

\usepackage{enumerate}
\usepackage{tensor,multirow}
\usepackage{upgreek}
\usepackage{url}
\usepackage{float}
\usepackage[pdfencoding=auto, psdextra,linktocpage]{hyperref}
\definecolor{darkblue}{rgb}{0, 0.3, 0.6}
\hypersetup{
    colorlinks=true,
    linkcolor=darkblue,
    filecolor=darkblue,      
    urlcolor=darkblue,
    citecolor = darkblue,
    pdftitle={Particles with precessing spin in Kerr spacetime: analytic solutions for eccentric orbits and homoclinic motion near the equatorial plane},
    }
\usepackage{orcidlink}

\newcommand{\FC}{\text{FC}}
\newcommand{\FT}{\text{FT}}
\newcommand{\FE}{\text{FE}}
\newcommand{\per}{\mathsf{per}}
\renewcommand{\hom}{\mathsf{hom}}
\renewcommand{\Im}{\mathrm{Im}\,} 

\newcommand{\dd}{\mathrm{d}}

\newcommand{\totder}[2]{\frac{\mathrm{d} #1}{\mathrm{d} #2}}

\newcommand{\Lzrd}{L_{z\text{rd}}}

\DeclareMathOperator\arctanh{tanh^{-1}}

\newcommand{\bea}{\begin{eqnarray}}
\newcommand{\eea}{\end{eqnarray}}
\newcommand{\be}{\begin{equation}}
\newcommand{\ee}{\end{equation}}
\newcommand{\ba}{\begin{align}}
\newcommand{\ea}{\end{align}}
\newcommand{\sgn}{\mathrm{sgn}}

\begin{document}
\title{
Particles with precessing spin in Kerr spacetime: analytic solutions for eccentric orbits and homoclinic motion near the equatorial plane 
}
\author{
Gabriel Andres Piovano$^1$ \orcidlink{0000-0003-1782-6813}
}
\email{gabrielandres.piovano@umons.ac.be}
\affiliation{Université Libre de Bruxelles, BLU-ULB Brussels Laboratory of the Universe, International
Solvay Institutes, CP 231, B-1050 Brussels, Belgium}
\affiliation{
Physique de l’Univers, Champs et Gravitation, Universit\'{e} de Mons – UMONS,
Place du Parc 20, 7000 Mons, Belgium
}%

\begin{abstract} 
We present a family of analytic solutions for the nearly-equatorial motion of a test particle with precessing spin in Kerr spacetime. We solve the equations of motion up to linear order in the small body's spin for periodic and homoclinic orbits. At zero order, the particle moves along equatorial geodesics. The spin-curvature force introduces post-geodesic corrections which, for generic spin orientations, cause the precession of the orbital plane. We derive the solutions for eccentric orbits in terms of Legendre elliptic integrals and Jacobi elliptic functions for generic referential geodesics (known as ``spin gauges"). Our analytical solutions perfectly match the numerical trajectories obtained by Drummond and Hughes in Phys. Rev. D 105, 124041 (2022), and Piovano et al. in Phys. Rev. D 111, 044009 (2025). Furthermore, we present, for the first time, the solutions for homoclinic orbits for a spinning particle in Kerr spacetime, and the spin-corrections to the location of the separatrix. The homoclinic trajectories are described in closed form using elementary functions. Finally, we introduce a novel parametrization for the motion of a spinning particle, called ``fixed eccentricity spin gauge". This is the only spin gauge in which the corrections to periodic orbits are finite at the geodesic separatrix, and continuously reduce to the last stable orbits under appropriate limits. Our results will be useful for modeling the inspiral and transition-to-plunge phases of asymmetric mass binaries within the two-time-scale framework.
\end{abstract}
\maketitle

\section{Introduction}
Upcoming space-based detectors such as LISA~\cite{LISA:2017pwj,LISA:2024hlh}, TianQin~\cite{TianQin:2015yph}, and Tajii~\cite{Ruan:2020smc} will be sensitive to gravitational wave (GW) sources that are unobservable by current terrestrial observatories. Prime examples are extreme-mass-ratio inspirals (EMRIs), astrophysical binaries formed by the capture of a stellar mass compact object by a supermassive black hole (SMBH) with mass $M \sim 10^5 - 10^7 M_\odot$. Due to their small mass ratios $q = \mu/M$, the smaller body (the secondary) completes $1/q$ orbits deep in the strong field of its larger companion (the primary), resulting in rich GW signals. Moreover, an EMRI signal can last for years in the detector band. For these reasons, the properties of the binary and its surrounding environment could be measured with excellent precision~\cite{Barack:2003fp,Babak:2017tow,Katz:2021yft,Speri:2023jte,Chapman-Bird:2025xtd}, leading to stringent tests of general relativity~\cite{Barack:2006pq,Gair:2012nm,Speri:2024qak,LISA:2022kgy}.

EMRI systems can be accurately modeled using black hole perturbation theory methods~\cite{LISAConsortiumWaveformWorkingGroup:2023arg}. Such approaches leverage the large disparity in the EMRI masses by iteratively solving the dynamics with a perturbative expansion in $q\ll 1$. At zeroth order, the smaller body moves along the geodesics of the primary spacetime. Post-geodesic corrections arise from the spin and higher multipoles of the compact body, and self-force (SF)~\cite{Pound:2021qin}. A two-timescale expansion reveals how the different post-geodesic terms affect the accuracy of EMRI waveforms~\cite{Hinderer:2008dm, Miller:2020bft}. The dominant contribution to the GW phase is given by the dissipative first-order in the mass ratio SF, which drives the adiabatic inspiral of the compact body~\cite{Hughes:2021exa}. Sub-leading corrections are labeled first post-adiabatic (or 1PA), and comprises the conservative first-order~\cite{vandeMeent:2017bcc,Nasipak:2025tby} and dissipative second-order in the mass ratio SF~\cite{Pound:2021qin,Wardell:2021fyy,Warburton:2021kwk}, and the secondary spin~\cite{Mathews:2025nyb}. Including all 1PA effects in EMRI models is crucial to avoid systematic biases in the recovery of the binary parameters~\cite{Burke:2023lno}. 

Based on these motivations, this paper investigates the dynamics of a spinning test particle in Kerr spacetime. Post-geodesic effects due to the compact body spin contribute to linear order in spin at the post-adiabatic phase. These include dissipative corrections to GW fluxes~\cite{Akcay:2019bvk,Grant:2024ivt,Piovano:2024yks,Skoupy:2023lih,Skoupy:2025nie} and conservative effects due to the spin-curvature force~\cite{Akcay:2015pza,Akcay:2016dku,Akcay:2017azq,Warburton:2017sxk,Drummond:2022_near_eq,Drummond:2022efc,Drummond:2023wqc,Witzany:2024ttz}. In this work, we focus on the latter and neglect radiation reaction and other SF effects. 

Great progress has been made in solving the equations of motion for generic, periodic orbits with numerical schemes. An efficient frequency domain method was first implemented in Refs.~\cite{Ruangsri:2015cvg,Drummond:2022efc,Drummond:2022_near_eq}, and later refined in~\cite{Skoupy:2023lih}, while a scheme based on osculating geodesics was proposed in~\cite{Drummond:2023wqc}. An alternative approach was put forward in~\cite{Witzany:2019nml}, which takes advantage of the complete integrability of the particle's motion in the linear regime~\cite{Rudiger:1981,Rudiger:1983,Kubiznak:2011ay}, (see also~\cite{Compere:2021kjz,Akpinar:2025tct,Ramond:2024ozy} for the quadratic and higher orders in the spin). Ref.~\cite{Witzany:2019nml} elegantly reduced the equations of motion to first order form using the Hamilton-Jacobi formalism and Hamiltonian of~\cite{Witzany:2018ahb}. Building on these results, Refs.~\cite{Piovano:2024yks,Skoupy:2025nie} implemented an efficient semi-analytic method to compute the orbits in the frequency domain, while~\cite{Witzany:2024ttz} found closed-form expressions for the actions and frequencies (see also~\cite{Gonzo:2024zxo} for a different derivation).

Finding analytic expressions for the spinning trajectories is challenging, since the equations of motion are not separable even in first order form~\cite{Witzany:2019nml} (with the exception of Schwarzshild~\cite{Witzany:2023bmq} and Reissner–Nordstr{\"o}m~\cite{Ciou:2025ygb} spacetime). Analytic solutions for spins aligned, equatorial orbits in Kerr spacetime were found in~\cite{Hackmann:2014tga} in terms of hyper-elliptic Kleinian functions, but these expressions are very cumbersome and difficult to manipulate (see e.g. Ref.~\cite{Hackmann:2008zz}).
Ref.~\cite{Skoupy:2024uan} achieved a major breakthrough by finding analytical solutions for generic orbits, which are given in terms of ``virtual geodesics" parametrized with a deformed Mino-time. However, the transformation between deformed and standard Mino-time is not separable. Thus, the solutions of~\cite{Skoupy:2024uan} can not be written in closed form as physical geodesics plus spin corrections in Mino-time. Moreover, it is not known how to map the solutions of~\cite{Skoupy:2024uan} to the numerical results of \cite{Skoupy:2022adh,Drummond:2022_near_eq,Drummond:2022efc,Piovano:2024yks}.

At the edge of the parameter space for periodic orbits lies the separatrix, which separates plunging and stable motion. The separatrix corresponds to a special class of geodesics called homoclinic~\cite{Levin:2008yp,Perez-Giz:2008ajn,Li:2023bgn,Ng:2025maa,Stein:2019buj}, with the innermost stable spherical~\cite{Teo:2020sey} and circular orbits~\cite{Bardeen:1972_ISCO} (ISSO and ISCO, respectively) representing degenerate cases. Several works have considered the spin corrections to the ISCO~\cite{Jefremov:2015gza,Suzuki:1997by}, and ISSO~\cite{Mukherjee:2018bsn,Skoupy:2025nie}. However, homoclinic motion for spinning particles was only studied in Ref.~\cite{Ciou:2025ygb} for Reissner–Nordstr{\"o}m spacetime, while the location of the separatrix was determined numerically in Ref.~\cite{Skoupy:2021asz} for equatorial orbits including high order terms in the spin. 
Exact solutions for the spin-corrections to homoclinic orbits could be useful in studying the transition from inspiral to plunge, which is particularly relevant in modeling Intermediate Mass Ratio Inspirals (IMRIs) or less asymmetric binaries~\cite{OShaughnessy:2002tbu,Sperhake:2007gu,Leather:2025nhu,Nagni:2025cdw,Lhost:2024jmw,Compere:2021zfj,Kuchler:2024esj}.

With these motivations in mind, this paper investigates the motion of a test particle with precessing spin near the equatorial plane of a Kerr spacetime. The misalignment of the binary spins induces the small precession of the orbital angular momentum, and, in turn, of the orbital plane. The equations of motion are solved by quadrature at linear order in the spin for eccentric and homoclinic orbits. In both cases, we found analytic solutions that are completely separated into physical geodesics plus subleading spin corrections. Periodic orbits are described by Legendre elliptic integrals and Jacobi elliptic functions for generic spin-gauges~\cite{Mathews:2025nyb,Piovano:2024yks}. For specific choices of the referential geodesic, we found the analytic counterparts of the numerical trajectories of Refs.~\cite{Drummond:2022_near_eq,Piovano:2024yks}. Moreover, we proved that the spin-corrections to the orbits diverge at the geodesic separatrix in all but one spin gauge, called the ``fixed eccentricity gauge". Such a gauge allowed us to compute, for the first time, the homoclinic orbits for a spinning particle in Kerr spacetime. The linear-in-spin correction to the trajectory and the location of the separatrix are given in closed form.

This paper is organized as follows. Section~\ref{sec:spinning_motion} first gives a review of the motion of a spinning particle in Kerr spacetime. Then, it introduces the equations of motion for nearly equatorial orbits. In Sec.~\ref{sec:equatorial_orbits} and Sec.~\ref{sec:polar_motion_spin_prec}, we present the solutions by quadrature for the equatorial and polar motion, respectively. Finally, Sec.~\ref{sec:discussion} summarizes the result of this work and gives possible extensions.

\subsection{Notation}
This work uses geometrized units $G=c=1$. Moreover, the mass $M$ of the primary is set to $M=1$. All spacetime indices are denoted by Greek letters, while tetrad legs are denoted with uppercase Latin letters. Partial and covariant derivatives are indicated with a comma and semicolon, respectively, i.e. $f_{\mu,\nu}= \partial_\nu f_\mu$ and  $f_{\mu;\nu}= \nabla_\nu f_\mu$. The metric signature is $(-,+,+,+)$, with the Riemann tensor defined as 
\begin{equation}
 \tensor{R}{^\delta_\sigma_\mu_\nu} \omega_\delta = 2\nabla_{[\mu} \nabla_{\nu]} \omega_\sigma
\end{equation}
where $\omega_\delta$ an arbitrary 1-form, while the square brackets denote antisymmetrization.
$\epsilon_{\alpha\beta\gamma\delta} = \sqrt{-g} e_{\alpha\beta\gamma\delta}$ the antisymmetric Levi-Civita tensor density and $e_{\alpha\beta\gamma \delta}$ the Levi-Civita symbol ($e_{0123} =1$).

\section{Equations of motion for a spinning particle}\label{sec:spinning_motion}
In this Section, we briefly review the dynamics of an extended body in curved spacetime. First, we touch upon the salient features of the motion of a spinning test particle in Kerr spacetime. Then, we introduce in Sec.~\ref{sec:EoM_near_eq} the equations of motion for the special case of nearly equatorial orbits. 

The Kerr geometry is described in Boyer-Lindquist coordinates by the line element
\begin{align}
   \dd s^2 &= -\left(1 - \frac{2 r}{\Sigma}  \right) \dd t^2 + \frac{\Sigma}{\Delta} \dd r^2 + \frac{\Sigma}{1-z^2} \dd z^2 + \nonumber \\ 
   &+\frac{1-z^2}{\Sigma}\left[2 a^2 r (1-z^2)+(a^2+r^2)\Sigma\right] \dd \phi^2 - \nonumber\\
   &- \frac{4 a r (1-z^2)}{\Sigma} \dd t \dd \phi \, ,
\end{align}
with  $z = \cos(\theta)$, $\Delta = r^2 -2 r + a^2$ and $\Sigma = r^2 + a^2 z^2$ while $a$ is the SMBH spin, which is aligned to the z-axis of a Cartesian coordinate system centered on the primary, namely $a \geq 0$. 

Under appropriate conditions, a small body moving in a curved spacetime is well described by the motion of a test particle endowed with small corrections, known as multipoles. Such corrections can be obtained through a suitable expansion of body's stress-energy tensor $T^{\mu \nu}$. In the pole-dipole approximation of celestial bodies, only the first two multipoles are retained, namely the mass $\mu$ and intrinsic angular momentum $S$ (or simply spin)
\begin{equation}
    \mu^2 = -p^\sigma p_\sigma \qquad S=\frac{1}{2} S^{\mu \nu}S_{\mu\nu}
\end{equation}
with $p^\mu$ and $S^{\mu \nu}$ being, respectively, the four-momentum and anti-symmetric spin tensor of the body. A spin vector can then be defined used the aforementioned quantities as
\begin{equation}
    S^\mu = - \frac{1}{2 \mu} \epsilon^{\mu\nu\rho\sigma} p_\nu S_{\rho\sigma} \,. \label{eq:spinvectordef}
\end{equation}
Thus, at pole-dipole order, the small compact object is approximated as a spinning test-particle, whose dynamics is governed by the Mathisson-Papapetrou-Dixon equations \cite{Mathisson:2010, Papapetrou:1951pa, Dixon:1970I}
\begin{subequations}
\label{eq:MPDeq}
  \begin{empheq}{align}
     \frac{D p^\mu}{\dd \tau} &= -\frac{1}{2} R^{\mu}_{~\nu \rho \sigma} v^\nu S^{\rho \sigma}  \, ,  \label{eq:spin_curvature} \\ 
     \frac{D S^{\mu\nu}}{\dd \tau} &= 2 p^{[\mu}v^{\nu]}   \, , 
  \end{empheq}
\end{subequations}
where $D/\dd \tau = v^\mu \nabla_\mu$,$\tau$ is the proper time, $R^{\mu}{}_{\nu\kappa\lambda}$ is the Riemann tensor, $x^\mu(\tau)$ is the representative worldline of the body and $v^\mu$ is its tangent vector. The right-hand side of Eq.~\eqref{eq:spin_curvature} is known as the spin-curvature force, which sources the leading post-geodesic effect due to the small body spin. 

The MPD equations do not constitute a closed system, since there exists an intrinsic freedom to choose the centroid of an extended body. To uniquely fix a referential wordline, the MPD equations must be supplemented with a spin condition. A detail discussion of the most commonly employed supplementary spin conditions is given in~\cite{Semerak:1999qc}. This work adopts the convenient Tulczyjew-Dixon condition \cite{tulczyjew1959motion,Dixon:1970I}:
\begin{equation}
S^{\mu\nu}p_\nu =0 \, ,
\end{equation}
which ensures that the mass $\mu$ and spin magnitude $S$ are first integrals of motion. Moreover, the specific spin vector $s^\rho =S^\rho/\mu$ can be expressed in terms of the specific spin tensor $s^{\rho\sigma} = S^{\rho \sigma}/\mu$ as
\begin{equation}
    s^{\mu\nu} = \frac{1}{\mu}\epsilon^{\mu\nu\alpha\beta}p_\alpha s_{\beta} \, . \label{eq:spintensordef}
\end{equation}
where $u^\sigma = p^\sigma/\mu$. It was shown by Dixon~\cite{Dixon:1970I} that the MPD equations admit a first integral of motion $C_\kappa$ for any Killing vector $\kappa^\mu$ of the background geometry
\begin{equation}
    C_\kappa = p_\mu k^\mu - \frac{1}{2} \kappa^{(\phi)}_{\mu;\nu} s^{\mu\nu} \ .
\end{equation}
Since the Kerr geometry admits the Killing vectors $\kappa_{(t)}^\mu \partial_\mu = \partial_t$ and $\kappa_{(\phi)}^\mu \partial_\mu = \partial_\phi$, the following quantities
\begin{subequations}
\begin{align}
    E = - u_\mu \kappa_{(t)}^\mu + \frac{1}{2} \kappa^{(t)}_{\mu;\nu} s^{\mu\nu} \,, \\
    J_z = u_\mu \kappa_{(\phi)}^\mu - \frac{1}{2} \kappa^{(\phi)}_{\mu;\nu} s^{\mu\nu} \,,
\end{align}
\end{subequations}
are conserved. At infinity, $E$ and $J_z$ can be interpreted, respectively, as the orbital energy (normalized by $\mu$) and total orbital angular momentum parallel to the z-axis (rescaled by $\mu M$) of the compact body.
  
Eqs.~\eqref{eq:MPDeq} form a non-linear, coupled system of differential equations, and therefore can not be solved in closed form for generic orbital configurations. 
However, Eqs.~\eqref{eq:MPDeq} can be simplified by noticing that the spin magnitude $S$ is small in most astrophysical scenarios, as observed in~\cite{Hartl:2002ig}. Indeed, the spin magnitude $S$ has the same dimension of an angular momentum in geometrized units
\begin{equation}
\frac{S}{\mu M} = q \chi \,, \quad \chi \equiv S/ \mu^2 \ ,
\end{equation} 
where the mass $M$ was restored for clarity, while $\chi$ is the dimensionless spin of the test body. The secondary spin is expected to be $|\chi| \leq 1$ for EMRIs and IMRIs observable by future space observatories, and for known sources detectable by terrestrial observatories\footnote{Neutron stars and black holes are the only known compact objects that can spiral deep into the strong field of the SMBH without being tidally disrupted. For neutron stars, $|\chi| \sim \mathcal O(1/10)$ due to mass shed limits~\cite{Hartl:2002ig}, whereas the Kerr bound imposes that $|\chi| \leq 1$ for black holes. Moreover, all observations made so far by the LIGO and Virgo detectors are compatible with astrophysical binaries of black holes, neutron stars or neutron star-black hole~\cite{LIGOScientific:2025slb}.}. Thus, the spin magnitude $S$ scales as the mass ratio, thereby $S \ll 1$. 

Waveform models accurate up to 1PA order only require solutions of the MPD equations accurate up to order $\mathcal O(q \chi)$~\cite{Mathews:2025nyb}. The velocity $v^\mu$ and linear momentum $p^\mu$ are colinear at order $\mathcal O(q \chi)$, $u^\mu = v^\mu + \mathcal O(q^2 \chi^2)$~\cite{Semerak:1999qc}, thereby the MPD equations~\eqref{eq:MPDeq} at linear order in $q \chi$ reduces to 
\begin{subequations}
\label{eq:MPDeq_lin}
  \begin{empheq}{align}
    \frac{D_{\rm g} v^\mu_{\rm g}}{\dd \tau} &= 0   \, , \label{eq:2ndEoMgeo}  \\
    \frac{D_{\rm g} \delta v^\mu}{\dd \tau} &  = -\frac{1}{2} R^{\mu}_{~\nu \rho \sigma} v^\nu_{\rm g} s^{\rho \sigma} \, , \label{eq:2nd_order_EoM_linear_momentum} \\
    \frac{D_{\rm g} s^{\mu}}{\dd \tau} &= 0 \, ,  \label{eq:parallel_transport_spin_vector}  
  \end{empheq}
\end{subequations}
with $v^\mu_{\rm g}$ the geodesic 4-velocity and $\delta v^\mu$ the shifts due to the secondary spin, while $D_g/\dd \tau = v^\mu_{\rm g} \nabla_\mu$. We also used the fact that Eq.~\eqref{eq:parallel_transport_spin_vector} is equivalent to $D s^{\mu \nu} /\dd \tau = 0$ thanks to Eq.~\eqref{eq:spinvectordef}. 
Conveniently, R\"{u}dinger~\cite{Rudiger:1981,Rudiger:1983} showed that the system~\eqref{eq:MPDeq_lin} admits two approximate first-integrals, 
\begin{align}
    s_\parallel &= \frac{Y_{\mu\nu} u^\mu s^\nu}{\sqrt{K_{\mu\nu} u^\mu u^\nu}} \,, \\
    K &= K_{\mu\nu} u^\mu u^\nu + 4 u^\mu s^{\rho\sigma} Y^\kappa{}_{\left[\mu\right.} Y_{\left.\sigma\right]\rho;\kappa} \,,
\end{align}
which are conserved up to $\mathcal O(q \chi)$ order for a generic test-body. When the secondary is a Kerr BH, a modified version of the R\"{u}dinger constants are conserved up to quadratic order in the spin~\cite{Compere:2023alp}. These approximate constants of motion arise from an ``hidden symmetry" for the Kerr geometry, encoded by the existence of a Killing-Yano tensor $Y_{\mu\nu}$, $Y_{\mu(\nu;\kappa)} = 0$, with associated Killing tensor $K_{\mu\nu}$. The R\"{u}dinger constants $K$ and $s_\parallel$ can be interpreted, respectively, as a generalization of the Carter constant and the component of the spin vector parallel to the orbital angular momentum. It is then convenient to define $\chi_\parallel = s_\parallel/\mu$ and $\chi_\perp = \pm\sqrt{\chi^2 - \chi_\parallel^2}$. 

The existence of the R\"{u}dinger constants implies that Eqs.~\eqref{eq:MPDeq_lin} are completely integrable in a Liouville sense~\cite{Kubiznak:2011ay,Witzany:2019nml}. In Ref.~\cite{Witzany:2019nml}, Witzany uses this fundamental property and the Hamilton-Jacobi formalism to reduce Eqs.~\eqref{eq:MPDeq_lin} into an equivalent system of first order differential equations, which are easier to solve. Unfortunately, these first-order equations of motion remain non-separable, like their second-order counterparts~\eqref{eq:MPDeq_lin}. However, in the case of nearly equatorial orbits, it was shown that Eqs.~\eqref{eq:MPDeq_lin} partially decoupled~\cite{Tanaka:1996ht,Drummond:2022_near_eq}. In the following, we take advantage of first-order system of equations derived in~\cite{Witzany:2019nml} and the results of~\cite{Drummond:2022_near_eq}.

\subsection{Geodesic motion on the equatorial plane} \label{sec:geodesic_equatorial_motion}
We briefly present here some of the key features of periodic and homoclinic geodesics on Kerr equatorial plane. The geodesic equations of motion are
\begin{subequations} \label{eq:1st-order-EoMgeo_rad}
    \begin{empheq}[]{align}
        \totder{r_{\rm g}}{\lambda} & = \pm \sqrt{R_{\rm g}(r_{\rm g})}  \, ,\\
        \totder{t_{\rm g}}{\lambda} &= T_{\rm g}(r_{\rm g}) \, , \\
        \totder{\phi_{\rm g}}{\lambda} &=  \Phi_{\rm g}(r_{\rm g})  \, ,
    \end{empheq}
\end{subequations}
where $\lambda$ is the Mino-time parameter defined as $\dd \tau = \Sigma \dd \lambda$, while
\begin{align}
    & R_{\rm g}(r_{\rm g}) = (E_{\rm g} r^2_{\rm g} - a \Lzrd)^2 - \Delta (r^2_{\rm g} + \Lzrd^2) \label{eq:1st-order-EoMgeo}  \, , \\
    & T_{\rm g}(r_{\rm g}) = \frac{1}{\Delta}\Big( E_{\rm g} r^4_{\rm g}  + E_{\rm g} a^2 r^2_{\rm g}  -2 a \Lzrd r_{\rm g} \Big)  \, ,\\
    & \Phi_{\rm g}(r_{\rm g}) = \frac{1}{\Delta}\Big( (a E_{\rm g} + \Lzrd)r^2_{\rm g} - 2 \Lzrd r_{\rm g}  \Big)  \, . 
\end{align}
The quantities $E_{\rm g}$ and $L_{z \rm g}$ are the geodesic constants of motion, with $\Lzrd = L_{z \rm g} - a E_{\rm g}$. In particular, $\Lzrd>0$ ($\Lzrd<0$) corresponds to prograde (retrograde) orbits. The special case $\Lzrd = 0$ is only satisfied for a special class of plunging geodesics~\cite{Compere:2021bkk}, which will not be considered here\footnote{$\Lzrd = 0$ implies that $r_{\rm g} =0$ is a double root for the radial geodesic potential $R_{\rm g}(r_{\rm g})$. Such condition is never verified for homoclinic or periodic motion.}.

Geodesic orbits are characterized by the roots of the radial function $R_{\rm g}(r_{\rm g})$, which is a quartic polynomial in the radius $r_{\rm g}$
\begin{equation}
    R_{\rm g}(r_{\rm g}) = (1 - E^2_{\rm g})(r_{1\rm g} - r_{\rm g})(r_{\rm g} - r_{2\rm g})(r_{\rm g} - r_{3\rm g})r_{\rm g}  \, . \label{eq:radial_potential_periodic}
\end{equation}
Periodic and homoclinic motion only occur when the radial potential has four reals roots , which can be ordered by magnitude as $0 < r_{3\rm g}  \leq r_{2\rm g} \leq r_{1\rm g}$. Using Vieta's formula,  the root $r_{3 \rm g}$ can be written as
\begin{equation}
    r_{3 \rm g} = \frac{2}{1-E^2_{\rm g}} - (r_{1 \rm g} + r_{2 \rm g}) \ .
\end{equation}
It is convenient to parametrize the geodesic turning points $r_{1 \rm g}$ and $r_{2 \rm g}$ in terms of the Keplerian like orbital elements $p_{\rm g}$ (semi-latus rectum) and $e_{\rm g}$ (eccentricity) as
\begin{align}
    r_{1 \rm g} = \frac{p_{\rm g}}{1 - e_{\rm g}}  \qquad  r_{2 \rm g} = \frac{p_{\rm g}}{1 + e_{\rm g}}
\end{align}

Henceforth, we consider as initial conditions $(r_{\rm g}(0), t(0), \phi(0)) = (r_{1 \rm g}, 0, 0)$ for $\lambda =0$.\footnote{Our convention differs from the one adopted by the Black Hole Perturbation Toolkit~\cite{BHPToolkit} and Ref.~\cite{vandeMeent:2019cam}, which consider $r_{\rm g}(0) = r_{2 \rm g}$.} When all roots of $R_{\rm g}(r_{\rm g})$ are distinct, the motion is periodic, and the particle librates in the radial direction with Mino-time frequency $\Upsilon_{r \rm g}$. The coordinate time and azimuthal motion are instead characterized by the frequencies $\Upsilon_{t \rm g}$, and $ \Upsilon_{\phi \rm g}$, respectively. Periodic geodesics can be described in analytic form using special functions~\cite{Fujita:2009bp,vandeMeent:2019cam,Hackmann:2010zz,Cieslik:2023qdc}, while the constants of motion can be written in terms of the turning points $r_{1 \rm g}$ and $r_{2 \rm g}$~\cite{Schmidt:2002qk}.

When two or more roots of $R_{\rm g}(r_{\rm g})$ merge, the type of motion change drastically. Stable circular orbits are admissible in the case $r_{1 \rm g} = r_{2 \rm g}$, while the innermost stable circular orbit (or ISCO) corresponds to the triple root $r_{1 \rm g} = r_{2 \rm g} = r_{3 \rm g}$. Homoclinic orbits occur when $r_{2 \rm g} = r_{3 \rm g}$ and the initial radius $r_{\rm g}(0)$ is $r_{2 \rm g} < r_{\rm g}(0) \leq r_{1 \rm g}$, while unstable circular orbits are located at the double root $r_{2\rm g} = r_{3 \rm g}$. Homoclinic geodesics represent the separatrix between stable, bound orbits and plunging trajectories~\cite{Levin:2008yp,Perez-Giz:2008ajn}. The semi-latus recta $p^*_{\rm g} = p^*_{\rm g}(a,e)$ for which $r_{2 \rm g} = r_{3 \rm g}$ denote the location of the separatrix, which is a surface parametrized by $a$ and $e_{\rm g}$~\cite{Stein:2019buj}. We employ the ``KerrGeodesic" package~\cite{BHPToolkitKerrGeodesics} of the Black Hole Perturbation Toolkit~\cite{BHPToolkit} to numerically calculate $p^*_{\rm g}$.

A particle on a homoclinic geodesic takes an infinite Mino (or proper) time to reach the periastron $r_{2 \rm g}$, which implies that $\Upsilon_{r \rm g} \to 0$ as $r_{3 \rm g} \to r_{2 \rm g}$. For this reason, the homoclinic trajectories $t(\lambda)$ and $\phi(\lambda)$ diverge logarithmically as $r_{\rm g} \to r_{2 \rm g}$. However, the frequencies $\Upsilon_{t \rm g}$ and $\Upsilon_{\phi \rm g}$ are finite at the last stable orbits. Finally, in the limit $r_{3 \rm g} \to r_{2 \rm g}$, the analytic expressions for periodic orbits reduce to homoclinic geodesics, which are described by elementary functions.

Appendix~\ref{app:geo_analytic} collects the analytic expressions for periodic and homoclinic motion previously derived in the literature.

\subsection{Equations of motion for nearly equatorial orbits}\label{sec:EoM_near_eq}
We seek a perturbative solution of the system~\eqref{eq:MPDeq_lin} in the form 
\begin{align}    
    x^\mu(\tau) &= x^\mu_{\rm g}(\tau) + q \chi \delta x^\mu(x^\mu_{\rm g}(\tau), s^\mu(\tau))   \label{eq:trajectory_expansion}  \ .
\end{align}
In particular, we focus on orbits $\mathcal O(q \chi)$ close to the equatorial plane, with the polar trajectory $z =z (\tau)$ of the test body given by
\begin{equation}
 z(\tau) = 0 + q \chi_\parallel \delta z_\parallel(\tau) + q \chi_\perp \delta z_\perp(\tau) \ , \label{eq:condition_almost_equatorial_orbits}
\end{equation}
where $\delta z_\parallel$ ($\delta z_\perp$) is the correction due the parallel (orthogonal) component of the secondary spin with respect to the primary spin.

As already observed in~\cite{Tanaka:1996ht,Drummond:2022_near_eq}, the nearly equatorial motion can  be described by the superposition of planar motion constrained into the equatorial plane, and small oscillations orthogonal to the plane. As a result, the orbital plane slowly precess due to the misalignment of the secondary spin with the orbital angular momentum.

\subsubsection{Motion on the equatorial plane}
We now introduce the first order differential equations that describe the equatorial motion of the spinning particle. By imposing the constraint~\eqref{eq:condition_almost_equatorial_orbits} on the equations of motion of Ref.~\cite{Witzany:2019nml}, we get
\begin{subequations}\label{eq:1st-order-EoM}
    \begin{empheq}[]{align}
        & \bigg(\totder{r}{\lambda} \bigg)^{\!\!2} = \frac{R_5(r)}{r} \label{eq:R_of_r_velocity}   \, ,\\
        &\totder{t}{\lambda} = T_{\rm g}(r) + q \chi_\parallel \delta T(r)  \label{eq:T_of_r_velocity} \, ,\\
        &\totder{\phi}{\lambda} = \Phi_{\rm g}(r) + q \chi_\parallel \delta \Phi(r)  \label{eq:Phi_of_r_velocity} \, , 
    \end{empheq}
\end{subequations} 
where 
\begin{align}
    & \delta T(r) = \frac{2}{\Delta}( a E_{\rm g} r - \Lzrd)  -\frac{\Lzrd}{r} + \displaystyle \sum^2_{i=1} \frac{\partial T_{\rm g}}{\partial C_{i \rm g}} \delta C_i  \, ,\\
    & \delta \Phi(r) = \frac{a}{r \Delta}( a E_{\rm g} r  - \Lzrd) - E_{\rm g} + \displaystyle \sum^2_{i=1} \frac{\partial \Phi_{\rm g}}{\partial C_{i \rm g}} \delta C_i   \, , 
\end{align}
with $(C_{1 \rm g}, C_{2 \rm g}) = (E_{\rm g}, L_{z \rm g})$. The spin corrections to the constants of motion are represented by $\delta C_i$, with $(\delta C_1, \delta C_2) = (\delta E, \delta L_z)$.
The radial potential $R_5(r)$ is a quintic polynomial 
\begin{equation}
   R_5(r) = \displaystyle \sum^5_{i=0} c_i r^i   \, ,  
\end{equation}
whose coefficients $c_i, \, i = 0, \dots ,5$ are defined as
\begin{align}
    c_5 &= E^2_{\rm g} - 1 + q \chi_\parallel 2 E_{\rm g} \delta E \ , \\
    c_4 &= 2 \ , \\
    c_3 &= a^2(E^2_{\rm g} - 1) - L^2_{z \rm g} \nonumber \\
    & + q\chi_\parallel 2 \big(a^2 E_{\rm g} \delta E + L_{z \rm g} (E_{\rm g} - \delta L_z )\big)    \ , \\
    c_2 &= 2 \Lzrd^2 - q \chi_\parallel 2 \Lzrd \big(3 E_{\rm g} + 2 a \delta E - 2 \delta L_z\big) \ ,   \\
    c_1 &=0 \ ,  \\
    c_0 &=  q \chi_\parallel 2 a \Lzrd^2 \ .
\end{align}
Eqs.~\eqref{eq:1st-order-EoM} are equivalent to the second order Eqs. (5.10) - (5.12) of Ref.~\cite{Drummond:2022_near_eq}. Moreover, the system~\eqref{eq:1st-order-EoM} is the linearized version of the first order equations of Sajio et al.~\cite{Saijo:1998mn}. To check this, one needs to first expand their Eqs.(2.19)-(2.25) at linear order in $s$ and then substituting $s \to q \chi_\parallel$.
The solution of the system~\eqref{eq:1st-order-EoM} is presented in Sec.~\ref{sec:equatorial_orbits}.

\subsubsection{Spin vector and precession phase}
In the linear-in-spin regime, the spin vector (and spin tensor) are simply parallel transported along the background geodesics, thereby their evolution do not depend on the corrections $\delta x^\mu$. A closed form solution for Eq.~\eqref{eq:parallel_transport_spin_vector} was found in Ref.~\cite{marck1983solution} in terms of a parallel transported tetrad $e_0^\mu$, $e_1^\mu$, $e_2^\mu$, $e_3^\mu$, called the Marck tetrad.
Here the Marck tetrad is defined according to the convention of Ref.~\cite{vandeMeent:2019cam}. Then, the spin vector can be decomposed as
\begin{equation}
    s^\mu = s_\parallel e_3^\mu + s_\perp (e_1^\mu \cos\psi_{\rm p}(\lambda) + e_2^\mu \sin\psi_{\rm p}(\lambda)) \, ,
\end{equation}
where $\psi_{\rm p}(\lambda)$ is the precession phase of the spin vector. $\psi_{\rm p}(\lambda)$ satisfies a separable equation of motion~\cite{marck1983solution}, solved by van de Meent for generic periodic orbits in~\cite{vandeMeent:2019cam}. In the limit of almost equatorial orbits, $\psi_{\rm p}(\lambda)$ satisfies the simpler equation
\begin{equation}
    \totder{\psi_p}{\lambda} = \Psi_r(r_{\rm g})  \, , \label{eq:spin-precession-angle}
\end{equation}
where 
\begin{equation}
    \Psi_r(r_{\rm g}) = |\Lzrd| \frac{(r^2_{\rm g}+a^2)E_{\rm g}-a L_{z \rm g}}{\Lzrd^2 + r^2_{\rm g}} + a \sgn(\Lzrd) \ ,
\end{equation}
A solution for Eq.~\eqref{eq:spin-precession-angle} for equatorial, periodic motion can be obtained from the analytic expression provided by van de Meent for generic periodic orbits~\cite{vandeMeent:2019cam}. A closed from expression for $\psi_p(\lambda)$ along homoclinic orbits is instead presented in Sec.~\ref{sec:polar_motion_spin_prec}. 

\subsubsection{Polar motion}
The polar motion is governed by Eq. (5.15) of Ref.~\cite{Drummond:2022_near_eq}. After some algebraic manipulations, it reduces to
\begin{equation}
\frac{\dd^2 \delta z}{\dd \lambda^2} + \Upsilon^2_{z\rm g}\delta z = - 3 \chi_\perp\frac{\cos\psi_{\rm p}}{r^2_{\rm g}}L_{z\text{rd}}\sqrt{L^2_{z\text{rd}} + r^2_{\rm g}} \ , \label{eq:polar_EoM}
\end{equation}
with $\Upsilon_{z \rm g} = \sqrt{a^2(1 - E^2_{\rm g}) + L^2_{z \rm g}}$ the frequency of harmonic oscillations near $z_{\rm g} =0$. Solutions for Eq.~\eqref{eq:polar_EoM} are presented in Sec.~\ref{sec:polar_motion_spin_prec}.

\section{Equatorial motion} \label{sec:equatorial_orbits}
\subsection{Roots of the radial potential}\label{sec:roots_radial_potential}
To find the solutions for the system~\eqref{eq:1st-order-EoM}, it is convenient to locate the roots of the function $R_5(r)$, which is a quintic polynomial. As such, the roots of $R_5(r)$ can not be found in closed form for generic values of its coefficients. However, we expect that the largest three roots of $R_5(r)$, namely $r_1, r_2, r_3$, are $\mathcal O (q \chi_\parallel)$ closed to the roots of the geodesic potential:
\begin{equation}
    r_i = r_{i \rm g} + q \chi_\parallel \delta r_i  \qquad i = 1 \dots 3 \ ,
\end{equation}
with $r_1 > r_2 \geq r_3$. The shifts to the turning points $\delta r_1$ and $\delta r_2$ can be found as follows. First, we expand the right hand side of Eq.~\eqref{eq:R_of_r_velocity} using Eq.~\eqref{eq:trajectory_expansion}
\begin{equation}
\frac{R_5(r)}{r} = R_{\rm g}(r_{\rm g}) + q \chi_\parallel \big[ \delta R(r_{\rm g}) +  R'_{\rm g}(r_{\rm g}) \delta r(r_{\rm g}) \big]  \, ,   
\end{equation}
where the prime denotes derivation with respect to $r_{\rm g}$, while
\begin{align}
&  \delta R(r_{\rm g}) = R_s(r_{\rm g}) + \displaystyle \sum^2_{i =1} \frac{\partial R_{\rm g}}{\partial C_{i \rm g}} \delta C_i  \ , \\
&   R_s(r_{\rm g}) = \frac{2 a \Lzrd^2}{r_{\rm g}} - 6 E_{\rm g} \Lzrd r_{\rm g} + 2 E_{\rm g} L_{z \rm g} r^2_{\rm g} \ .
\end{align}
Then, the shifts to the turning points $\delta r_1 = \delta r(r_{1\rm g})$ and $\delta r_2 = \delta r(r_{2\rm g})$
are determined by the constraints
\begin{align} \label{eq:constraints_shifts_constants_of_motion}   
    \delta R(r_{1\rm g}) = - \delta r_1 R'_{\rm g}(r_{1\rm g})  \ , \\
    \delta R(r_{2\rm g}) = - \delta r_2 R'_{\rm g}(r_{2\rm g})  \, ,  
\end{align}
once the the shifts to the constants of motion $\delta E$ and $\delta L_z$ are specified. The correction $\delta r_3$, and the roots $r_4, r_5$ can be obtained by using the Vieta's formulas, which relate the roots of a polynomial with its coefficients. In our case, it is sufficient to use the following Vieta's formulas\footnote{Alternatively, one can divide the polynomial $R_5(r_{\rm g})$ by $(r_1 - r)(r - r_2)$. Such polynomial division has zero remainder, thanks to the shifts $\delta E$ and the $\delta L_z$, and a cubic polynomial as quotient. Expanding the roots of this cubic polynomial in $q \chi_\parallel$ gives the same expressions for $\delta r_3$, $r_4$ and $r_5$ shown in the main text.}
\begin{align}
    -\frac{c_0}{c_5} &= r_1 r_2 r_3 r_4 r_5 \ ,\\
    -\frac{c_4}{c_5} &= r_1 + r_2 + r_3 + r_4 + r_5  \ ,\\
    0 &= (r_1 r_2 r_3 r_4) + (r_1 r_2 r_3 r_5) + (r_1 r_3 r_4 r_5) \nonumber \\
    & + (r_1 r_2 r_4 r_5) + (r_2r_3 r_4 r_5) \ , 
\end{align}
After expanding both sides of the former constraints in $q \chi$ and equating terms by terms, we obtain
\begin{align}
    \delta r_3 &=  \frac{4 E_{\rm g} \delta E}{(1 - E^2_{\rm g})^2} - \delta r_1 - \delta r_2 - 2\delta r_4 \ ,  \\
    \delta r_4 &= -\frac{a}{4 \Lzrd^2}\big(a^2(1 - E^2_{\rm g})+ L^2_{z \rm g} \big) \ , \\
    r_4 &= \sqrt{-q \chi_\parallel} \sqrt{a} + q \chi_\parallel \delta r_4 + \mathcal O(q^{3/2})  \, \\
    r_5 &= -\sqrt{-q \chi_\parallel} \sqrt{a} + q \chi_\parallel \delta r_4 + \mathcal O(q^{3/2}) \ .
\end{align}
$r_4$ and $r_5$ form a pair of complex conjugate roots when $\chi_\parallel >0$. It may seem surprising that the roots $r_4$ and $r_5$ includes a $\mathcal O(\sqrt{q})$ term. However, these roots are consistent with the perturbative expansion of the equations of motion, since $\mathcal O(\sqrt{q})$ terms do not appears in the spin corrections to the orbits. Moreover, by expanding the factors $(r - r_4)(r - r_5)$, we get
\begin{equation}
    (r - r_4)(r - r_5) = r^2 - q \chi_\parallel 2\delta r_4 - q \chi_\parallel a  + \mathcal O(q^2 \chi^2) 
\end{equation}
which is again consistent with the fact that $R_5(r) \to r R_g(r)$ for $\chi_\parallel \to 0$. Finally, the roots $r_4$ and $r_5$ vanishes for $a \to 0$, as expected (see Ref.~\cite{Witzany:2023bmq}).

\begin{widetext}
The polynomial $R_5(r)$ can then be factorized as
\begin{equation}
    R_5(r) = -c_5(r_{1 \rm g} + q \chi_\parallel \delta r_1 - r)(r - r_{2 \rm g} - q \chi_\parallel \delta r_2)(r - r_{3 \rm g} - q \chi_\parallel \delta r_3)(r - r_4)(r -r_5) \ . \label{eq:fac_quintic_polynomial}
\end{equation}
\end{widetext}
Homoclinic motion for a spinning body occurs when $\delta r_2 = \delta r_3$ at the geodesic separatrix $p_{\rm g} = p^*_{\rm g}$ (i.e when $r_{2\rm g} = r_{3 \rm g}$).

\subsection{Fixing the referential geodesics}\label{sec:spin_gauges}
So far, we have not specified yet the shifts to the constants of motion $\delta E$ and $\delta L_z$. As observed in Refs.~\cite{Piovano:2024yks,Mathews:2025nyb}, defining the constants of motion shifts is equivalent to selecting a referential geodesic for the spinning trajectories. Each referential geodesic can then be interpreted as a specific ``spin-gauge" of the phase space, which correspond to a different parametrization of the motion. We consider here two spin gauges previously employed in the literature: the fixed turning points and the fixed constants of motion parametrizations, which are denoted as ``FT'' and ``FC'', respectively. Moreover, we introduce a novel spin gauge, in which the referential geodesics has the same eccentricity of the spin-perturbed orbits. We refer to this parametrization with the shorthand notation ``FE''. Henceforth, we use the superscript FT, FC and FE to label the orbital elements calculated in one of three spin gauges.

\subsubsection{Fixed constants of motion}
One of the simplest choice of referential geodesics consists in imposing $\delta E^\FC = \delta L^\FC_z =0$. Such gauge was first introduced by Witzany in Ref.~\cite{Witzany:2019nml}, and it was later adopted in Ref.~\cite{Witzany:2024ttz} to derive the shifts to the coordinate-time frequencies and actions in analytic form. For eccentric, equatorial orbits, the shifts to the turning points are
\begin{align} 
    \delta r^\FC_1  = - \frac{R_s(r_{1\rm g})}{R'_{\rm g}(r_{1 \rm g})} \, , \qquad
    \delta r^\FC_2  = - \frac{R_s(r_{2\rm g})}{R'_{\rm g}(r_{2 \rm g})} \, ,  
\end{align}
which diverge in the limit of circular orbits, as was already observed in Refs.~\cite{Witzany:2019nml,Skoupy:2025nie}.

\subsubsection{Fixed turning points}
A second convenient spin gauge can be defined by picking a referential geodesic with the same turning points of the spinning trajectories. In other words, the shifts $\delta E^\FT$ and $\delta L^\FT_z$ are chosen by imposing $\delta r^\FT_1 = \delta r^\FT_2 = 0$. The expressions for $\delta E^\FT$ and $\delta L^\FT_z$ are rather long, and are presented in Appendix~\ref{app:spin_constants_of_motion_corr}. The FT gauge was first introduced by Skuop\'{y} and Lukes-Gerakopoulos~\cite{Skoupy:2022adh} for eccentric, equatorial orbits, and it was later extended to generic orbits by Drummond and Hughes~\cite{Drummond:2022efc,Drummond:2022_near_eq}. As shown in Sec.~\ref{sec:periodic_motion}, a major problem of this parametrization is the divergence of the spin corrections to the orbits near the separatrix. 

\subsubsection{Fixed eccentricity}
We present here a novel spin gauge, which is defined by fixing the spin corrections to the constants of motion as
\begin{align}
    \delta E^\FE &= \delta E^\FT +  \frac{\partial E_{\rm g}}{\partial p} \delta p  \ , \\
    \delta L_z^\FE &= \delta L_z^\FT +  \frac{\partial L_{z\rm g}}{\partial p} \delta p  \ ,
\end{align}
with associated shifts to the turning points
\begin{align}
    \delta r_1^\FE = \frac{\delta p}{1 - e_{\rm g}}  \ , \quad
    \delta r_2^\FE = \frac{\delta p}{1 + e_{\rm g}}  \ . \label{eq:shift_turning_points_FE}
\end{align}
The shift $\delta p$ is defined by imposing that the spin corrections to the roots $\delta r_2$ and $\delta r_3$ coincide:
\begin{equation}
    \delta r^\FE_2 = \delta r^\FE_3 \ ,
\end{equation}
which gives
\begin{equation}
    \delta p = \frac{2(1 - e^2_{\rm g})\big(2 E_{\rm g} \delta E^\FT - \big(1 - E^2_{\rm g}\big)^2 \delta r_4\big)}{4 E_{\rm g} \frac{\partial E_{\rm g}}{\partial p} \big(e^2_{\rm g} -1 \big) + (3 - e_{\rm g}) \big(1 - E^2_{\rm g}\big)^2}  \ . \label{eq:semi_latus_rectum_shift}
\end{equation}
Eq.~\eqref{eq:shift_turning_points_FE} implies that the referential geodesic has the same eccentricity of the spinning trajectory. 

\subsection{Solutions by quadrature}\label{sec:quad_sol}
We now have all the necessary ingredients to find the spin corrections to the motion. Using Eq.~\eqref{eq:trajectory_expansion} and Eq.~\eqref{eq:fac_quintic_polynomial}, we expand the system~\eqref{eq:1st-order-EoM} to get the linear in spin corrections to the 4-velocities, which are given by
\begin{subequations}\label{eq:1st-order-EoM-lambda-fully-linearize}
    \begin{empheq}[]{align}
        \totder{\delta r}{\lambda} & = \pm \sqrt{R_{\rm g}(r_{\rm g})}\delta V_r(r_{\rm g}) \pm  \frac{\delta r (r_{\rm g})}{2\sqrt{R_{\rm g}(r_{\rm g})}} \totder{R_{\rm g}}{r_{\rm g}}  \, , \\
        \totder{\delta t}{\lambda} &= \delta T(r_{\rm g}) + \totder{T_{\rm g}(r_{\rm g})}{r_{\rm g}} \delta r(r_{\rm g}) \, , \\
        \totder{\delta \phi}{\lambda} &= \delta \Phi(r_{\rm g}) + \totder{\Phi_{\rm g}(r_{\rm g})}{r_{\rm g}} \delta r(r_{\rm g})  \, ,
  \end{empheq}
\end{subequations} 
where the upper (lower) sign correspond to positive (negative) geodesic radial velocity, while the function $\delta V_r(r_{\rm g})$ is defined as
\begin{align}
  \delta V_r(r_{\rm g}) &= -\frac{1}{2}\bigg( \displaystyle \sum^3_{i =1} \frac{\delta r_i}{r_{\rm g} - r_{i \rm g}} + \frac{2\delta r_4}{r_{\rm g}} - \frac{a}{r^2_{\rm g}} + \frac{2 E_{\rm g} \delta E}{1 - E^2_{\rm g}} \bigg)  \, .  \label{eq:diff_radial_potential} 
\end{align}
By changing variable from $\lambda$ to $r_{\rm g}$, the system~\eqref{eq:1st-order-EoM-lambda-fully-linearize} can be written as
\begin{subequations}\label{eq:1st-order-EoM-r-fully-linearize}
   \begin{empheq}[]{align}
     &\totder{\delta r}{r_{\rm g}} = \frac{1}{2 R_{\rm g}(r_{\rm g})} \totder{ R_{\rm g}}{r_{\rm g}}\delta r (r_{\rm g}) +\delta V_r(r_{\rm g}) \, ,  \label{eq:lin-corr-radial-motion}  \\  
	&\totder{\delta t}{r_{\rm g}} = \pm \frac{\delta T(r_{\rm g})}{\sqrt{R_{\rm g}(r_{\rm g})}} \pm \totder{T_{\rm g}(r_{\rm g})}{r_{\rm g}} \frac{\delta r(r_{\rm g})}{\sqrt{R_{\rm g}(r_{\rm g})}}  \, , \\ 
	&\totder{\delta \phi}{r_{\rm g}} = \pm \frac{\delta \Phi(r_{\rm g})}{\sqrt{R_{\rm g}(r_{\rm g})}} \pm \totder{\Phi_{\rm g}(r_{\rm g})}{r_{\rm g}}\frac{\delta r(r_{\rm g})}{\sqrt{R_{\rm g}(r_{\rm g})}}  \, .
   \end{empheq}
\end{subequations} 
Eq.~\eqref{eq:lin-corr-radial-motion} is a linear, inhomogeneous, first order differential equation, therefore its solution is given by
\begin{equation}
    \delta r(r_{\rm g}) = \sqrt{R_{\rm g}(r_{\rm g})} \int^{r_{1 \rm g}}_{r_{\rm g}} \frac{\delta V_r(r') \dd r'}{\sqrt{R_{\rm g}(r')}} \,  \label{eq:generic_solution_radial_shift}     
\end{equation}
Using Eq.~\eqref{eq:generic_solution_radial_shift} and integration by parts, the solutions $\delta t(r_{\rm g})$ and $\delta \phi(r_{\rm g})$ can be written as
\begin{align}
       \delta t(r_{\rm g}) &= \mp T_{\rm g}(r_{\rm g})\frac{\delta r(r_{\rm g})}{\sqrt{R_{\rm g}(r_{\rm g})}} \pm \int^{r_{1\rm g}}_{r_{\rm g}}\frac{\delta T(r')\dd r'}{\sqrt{R_{\rm g}(r')}}  \nonumber \\ 
       &\mp \int^{r_{1\rm g}}_{r_{\rm g}}\frac{T_{\rm g}(r') \delta V_r(r')}{\sqrt{R_{\rm g}(r')}} \dd r'  \,  \label{eq:generic_solution_time_shift}  \\
        \delta \phi(r_{\rm g}) &= \mp \Phi_{\rm g}(r_{\rm g})\frac{\delta r(r_{\rm g})}{\sqrt{R_{\rm g}(r_{\rm g})}} \pm \int^{r_{1\rm g}}_{r_{\rm g}}\frac{\delta \Phi(r')\dd r'}{\sqrt{R_{\rm g}(r')}}  \nonumber \\ 
       & \mp \int^{r_{1\rm g}}_{r_{\rm g}}\frac{\Phi_{\rm g}(r') \delta V_r(r')}{\sqrt{R_{\rm g}(r')}} \dd r'  \, . \label{eq:generic_solution_azimuthal_shift} 
\end{align} 
Notice that the right hand side of Eqs.~\eqref{eq:1st-order-EoM-r-fully-linearize} is not defined at the geodesic turning points $r_{1 \rm g}$ and $r_{2 \rm g}$. Thus, the solutions~\eqref{eq:generic_solution_radial_shift}-~\eqref{eq:generic_solution_azimuthal_shift} are also not defined at such points. However, these singularities are always removable for periodic orbits, since $R_{\rm g}(r_{\rm g})$ has not repeated roots. In this case, the limits $r_{\rm g} \to r_{2 \rm g}$ and $r_{\rm g} \to r_{1 \rm g}$ exist and are finite for the system~\eqref{eq:1st-order-EoM-r-fully-linearize}. As a result, the shifts to the velocities~\eqref{eq:1st-order-EoM-r-fully-linearize} and trajectories~\eqref{eq:generic_solution_radial_shift}-~\eqref{eq:generic_solution_azimuthal_shift} can be made differentiable at the geodesic turning points if they are suitably defined as piecewise functions.\footnote{An analogous case occurs for the function $\text{sinc}(x) =\sin(x)/x$, which is formally not defined for $x = 0$. However, the function can be made everywhere continuous (and even analytical) if it is defined piecewise as $\text{sinc}(x) =\sin(x)/x$ for $x \neq 0$ and $\text{sinc}(0)=1$ for $ x = 0$.} The only exception is the FT gauge, in which the shifts to the trajectories and velocities can be represented as not piecewise functions.

In the case of homoclinic motion, $R_{\rm g}(r_{\rm g})$ has a double root $r_{2 \rm g} = r_{3 \rm g}$. As a result, the singularity at $r_{\rm g} = r_{2 \rm g}$ can not be removed for the spin corrections to the trajectories $\delta t(r_{\rm g})$ and $\delta \phi(r_{\rm g})$, which diverge logarithmically as the particle approaches the periastron. Such divergent behavior is expected, as the geodesic trajectories $t_{\rm g}(r_{\rm g})$ and $\phi_{\rm g}(r_{\rm g})$ also diverge logarithmically near the periastron~\cite{Levin:2008yp,Mummery:2023hlo}.
    
\subsection{Periodic motion}\label{sec:periodic_motion}
We present here the analytic spin corrections to periodic motion valid for a generic spin gauge. After some algebraic manipulations involving partial fraction decompositions, the shifts to the trajectories $\delta r(r_{\rm g})$ $\delta t(r_{\rm g})$ and $\delta \phi(r_{\rm g})$ can be finally written as
\begin{align} 
    \delta r(r_{\rm g}) &= \sqrt{R_g(r_{\rm g})}\bigg[ \bigg(\frac{E_{\rm g}\delta E}{1 - E^2_{\rm g}} + \frac{\delta \Upsilon_r}{\Upsilon_{r\rm g}}\bigg) I^\per(r_{\rm g})  \nonumber \\
        & + \displaystyle \sum^3_{i=1} I^{\per}_{1/(r - r_i)}(r_{\rm g})\frac{\delta r_i}{2} - \frac{a}{2} I^\per_{1/r^2}(r_{\rm g})  \nonumber \\
        & + \delta r_{4} I^\per_{1/r}(r_{\rm g})  \bigg]\ , \label{eq:radial_shift_trajectory_periodic}
\end{align}

\begin{figure*}[!bth]
    {\centering
        \includegraphics[width=\textwidth]{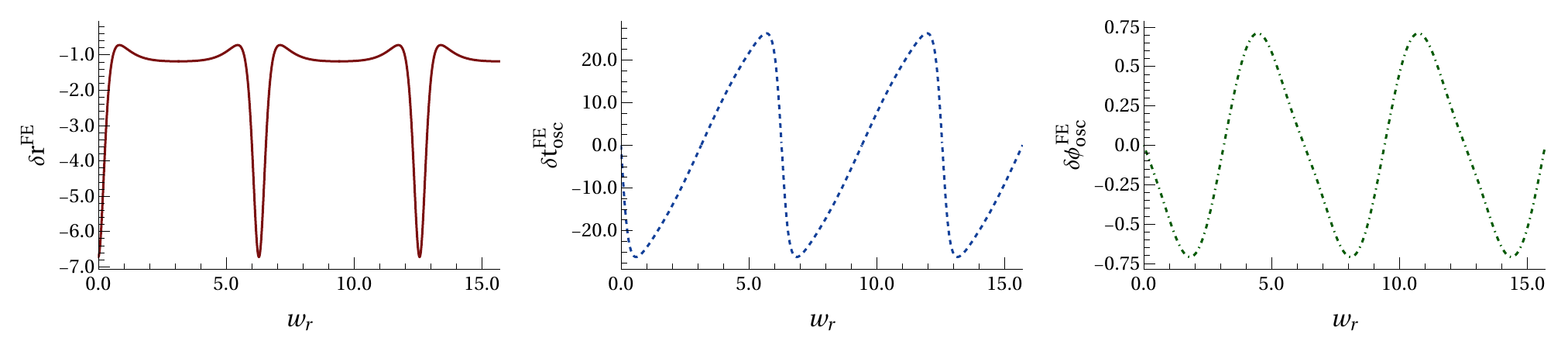}}
        \caption{Spin corrections to radial (left panel), coordinate time (middle panel) and azimuthal trajectories (right panel) for eccentric, prograde orbits for a spinning binary with $a =0.9$, $\chi_\parallel = 1/2$, $\chi_\perp = \sqrt{3}/2$. All shifts to the trajectories are computed in the FE gauge. Orbital parameters of the fiducial geodesic: $e_{\rm g} =0.7$, $p_{\rm g} = p^*_{\rm g} + 1/10$, with $p^*_{\rm g}$ the location of the geodesic separatrix at $e_{\rm g} = 0.7$.
}       \label{fig:shift_periodic_orbit}
\end{figure*}

\begin{widetext}
    \begin{align}
       \delta t(r_{\rm g}) &= - \frac{T_{\rm g}(r_{\rm g})\delta r(r_{\rm g})}{\sqrt{R_{\rm g}(r_{\rm g})}} + \displaystyle \sum^3_{i=1} T_{\rm g}(r_{i \rm g}) \frac{\delta r_i}{2} I^{\per}_{1/(r - r_i)}(r_{\rm g}) + \frac{\delta E}{1 - E^2_{\rm g}} \Big(I^\per_{r^2}(r_{\rm g}) + 2 I^\per_{r}(r_{\rm g}) + 4 I^\per(r_{\rm g}) \Big) \nonumber \\
        &  + \frac{E_{\rm g}}{2}\Big( \displaystyle \sum^3_{i=1} \delta r_i + 2 \delta r_4 \Big) I^\per_r(r_{\rm g}) + \frac{E_{\rm g}}{2} \Big(\displaystyle \sum^3_{i=1} (2 + r_{i \rm g})\delta r_i - a + 4\delta r_4 \Big) I^\per(r_{\rm g})   \ , \label{eq:time_shift_trajectory_periodic} 
    \end{align} 
\end{widetext}
\begin{align}
    \delta \phi(r_{\rm g}) &= - \frac{\Phi_{\rm g}(r_{\rm g})\delta r(r_{\rm g})}{\sqrt{R_{\rm g}(r_{\rm g})}} + \displaystyle \sum^3_{i=1} \Phi_{\rm g}(r_{i \rm g}) \frac{\delta r_i}{2} I^{\per}_{1/(r - r_i)}(r_{\rm g})  \nonumber \\
    & - \bigg(E_{\rm g} - L_{z \rm g}\frac{E_{\rm g} \delta E}{1 - E^2_{\rm g}} - \delta L_z \bigg) I^\per(r_{\rm g})   \ , \label{eq:azimuthal_shift_trajectory_periodic}
\end{align}
where the integrals $I^\per(r_{\rm g})$, $I^\per_{1/r}(r_{\rm g})$, $I^\per_{1/r^2}(r_{\rm g})$, $I^\per_{r}(r_{\rm g})$, $I^\per_{r^2}(r_{\rm g})$ and $I^\per_{1/(r - r_i)}(r_{\rm g})$ for $i= 1, \dots 3$ are given in Appendix~\ref{app:integrals_periodic_motion}. Moreover, we used the special functions described in Appendices~\ref{app:integrals_periodic_motion} and~\ref{app:geo_analytic} to parametrize the radial geodesic $r_{\rm g}(w_r)$ as a function of the relativistic mean anomaly $w_r = \big(\Upsilon_{r\rm g} + q \chi_\parallel \delta \Upsilon_r \big)\lambda$, with $\delta \Upsilon_r$ the spin corrections to the radial frequency. This parametrization is the same adopted in Refs.~\cite{Drummond:2022_near_eq,Drummond:2022efc,Skoupy:2023lih,Piovano:2024yks}. In particular, the term proportional to $\delta \Upsilon_r/\Upsilon_r$ in Eq.~\eqref{eq:radial_shift_trajectory_periodic} ensures that there are no secular terms growing with time in the shift to the radial motion, as apparent in the left panel of Fig.~\ref{fig:shift_periodic_orbit}.

The spin corrections to the radial frequency $\delta \Upsilon_r$ is formally defined by the integral
\begin{equation}
    \delta \Upsilon_r = \frac{\Upsilon^2_{r \rm g}}{\pi}\int^{r_{1 \rm g}}_{r_{2 \rm g}} \frac{\delta V_r(r_{\rm g}) \dd r'}{\sqrt{R_{\rm g}(r')}} \ , \label{eq:divergent_shift_radial_frequency}
\end{equation}
which is, unfortunately, divergent for all spin gauges but the FT gauge. However, the linear correction $ \delta \Upsilon_r$ is physically well defined and finite, as numerically shown by~\cite{Drummond:2022_near_eq,Skoupy:2022adh,Piovano:2024yks}. A physical result can be obtained for all spin gauges by defining $\delta \Upsilon_r$ as the \textrm{finite part} of the integral~\eqref{eq:divergent_shift_radial_frequency}, which can be obtained by means
of Hadamard's partie finie method~\cite{Blanchet:2000nu,Witzany:2024ttz}. The explicit expression of the regularized integral~\eqref{eq:divergent_shift_radial_frequency} is given in Appendix~\ref{app:spin_frequency_corr}. 
The shifts to the Mino-time frequencies for the coordinate time and azimuthal trajectories $\delta \Upsilon_t$ and $\delta \Upsilon_\phi$, respectively, can be written as
\begin{align}
    \delta \Upsilon_t &= \frac{\Upsilon_{r \rm g}}{\pi}\delta t(r_{2\rm g}) + \frac{\Upsilon_{t \rm g}}{\Upsilon_{r\rm g}}\delta \Upsilon_r \ , \\
    \delta \Upsilon_\phi &= \frac{\Upsilon_{r \rm g}}{\pi}\delta \phi(r_{2 \rm g}) + \frac{\Upsilon_{\phi \rm g}}{\Upsilon_{r\rm g}}\delta \Upsilon_r  \ ,
\end{align}
which do not require any regularization. Analytic expressions for $\delta \Upsilon_t$ and $\delta \Upsilon_\phi$ are given in Appendix~\ref{app:spin_frequency_corr}. The purely oscillatory component of the shifts $\delta t$ and $\delta \phi$ can then be expressed as
\begin{align}
    \delta b_\text{osc}(w_{\rm g})  &= \delta b(w_{\rm g}) - \frac{w_{r_{\rm g}}}{\Upsilon_{r\rm g}} \bigg(\delta \Upsilon_b - \frac{\Upsilon_{b \rm g}}{\Upsilon_{r\rm g}}\delta \Upsilon_r \bigg)   \ ,
\end{align}
with $b = (t,\phi)$. 

We checked the analytic expressions~\eqref{eq:radial_shift_trajectory_periodic}-\eqref{eq:azimuthal_shift_trajectory_periodic} in the FT gauge against the numerical results of Drummond and Hughes~\cite{Drummond:2022_near_eq}, and found perfect agreement. Moreover, Eqs.~\eqref{eq:radial_shift_trajectory_periodic}-\eqref{eq:azimuthal_shift_trajectory_periodic} completely match the numerical trajectories in the FT, FC and FE gauge derived using the code~\cite{repoHJ} developed in Ref.~\cite{Piovano:2024yks}. The code used for the calculation of the orbits, and all numerical checks are available in the GitHub repository~\cite{repoAnalyticHJproject}. 

\begin{figure}[!bth]
    {\centering
        \includegraphics[width=0.43\textwidth]{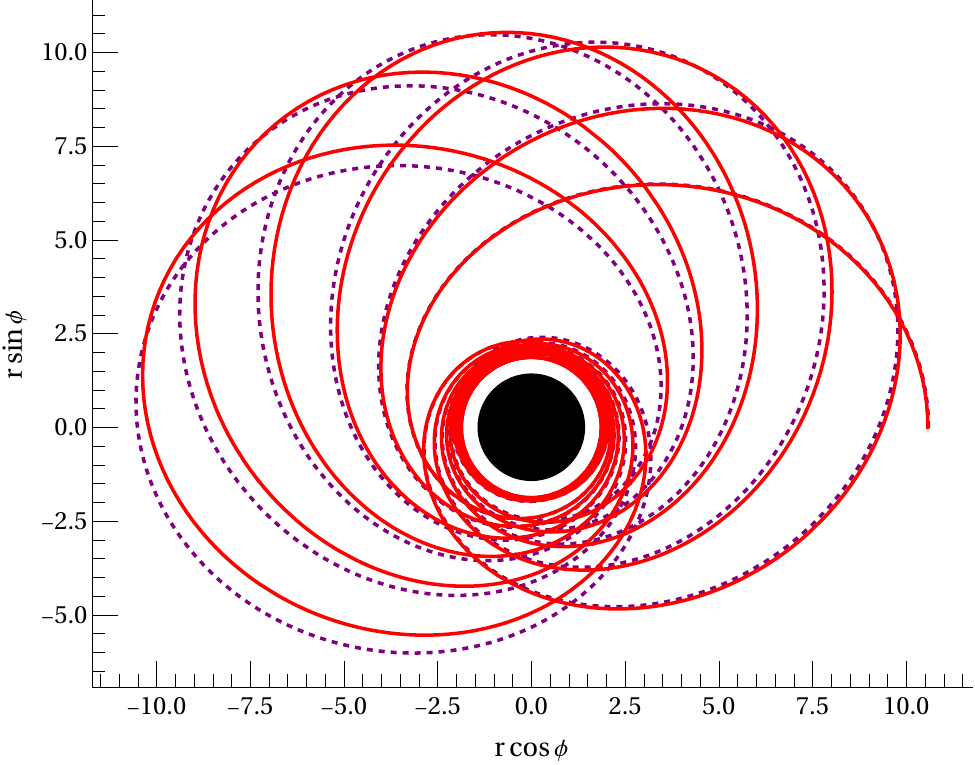}\\[0.4em]
        \includegraphics[width=0.43\textwidth]{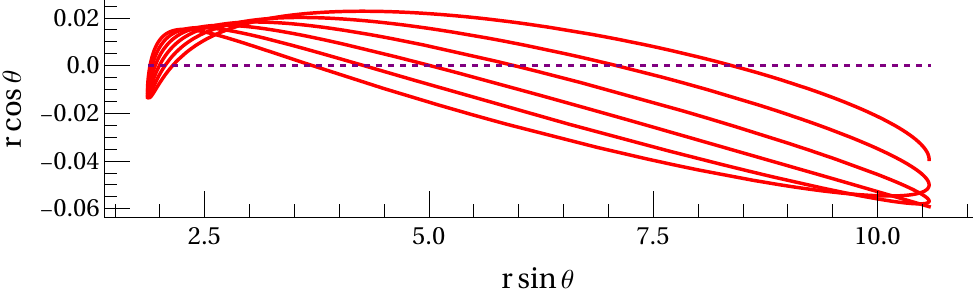}
    }
        \caption{Plots of zoom-whirl orbits~\cite{Glampedakis:2002ya} for a spinning particle (red-line) and its underlying fiducial geodesic (purple, dashed line). The orbital parameters and the linear-in-spin corrections are the same of Fig.~\eqref{fig:shift_periodic_orbit}, while the mass-ratio is fixed to $q = 1/100$. The radius of the black disc corresponds to the SMBH outer horizon $r_+ = 1+\sqrt{1 -a^2}$.
        Top panel:  projection of the orbits onto the equatorial plane. Bottom panel: orthogonal projection of the orbits onto a co-rotating polar plane.
}       \label{fig:zoom_whirl_orbit}
\end{figure}

Fig.~\ref{fig:shift_periodic_orbit} provides an example of the linear-in-spin shift to the trajectories in the FE gauge. An example of zoom-whirl orbit for a spinning and non-spinning particle is presented in the top panel of Fig.~\ref{fig:zoom_whirl_orbit} with secondary spin $(\chi_\parallel, \chi_\perp) = (1/2,3/2)$ and  for mass ratio $q = 10^{-2}$, which falls in the heavy IMRI domain~\cite{Bellovary:2025ris}. Close to the separatrix, the periastron advance for a spinning trajectory and its geodesic counterpart are remarkably different.

Thanks to Eq.~\eqref{eq:elliptic_amplitude_am} and the integrals of Appendix~\ref{app:integrals_periodic_motion}, it is easy to see that Eqs.~\eqref{eq:radial_shift_trajectory_periodic}-\eqref{eq:azimuthal_shift_trajectory_periodic} continuously reduce to stable, circular orbits in the limit $r_{2\rm} \to r_{1 \rm g}$ for the FC and FE gauge. By contrast, the solutions~\eqref{eq:radial_shift_trajectory_periodic}-\eqref{eq:azimuthal_shift_trajectory_periodic} diverge near the geodesic separatrix in all spin gauges but the FE gauge. Such singular behavior is due to $1/(r_{2 \rm g} - r_{3 \rm g})$ factors that appears in the integrals $I^\per_{1/(r - r_2)}(r_{\rm g})$ and $I^\per_{1/(r - r_3)}(r_{\rm g})$. The singular terms for $r_{3\rm g} \to r_{2 \rm g}$  cancels only when $\delta r_2 = \delta r_3$, a condition that is only satisfied in the FE gauge. Thus, the spin correction to the periodic orbits in the FE gauge continuously reduce to homoclinic orbits in the limit $r_{3 \rm g} \to r_{2 \rm g}$.

\subsection{Homoclinic motion and the location of the separatrix}\label{sec:homoclinic_motion}
\begin{figure}[!bth]
    {\centering
        \includegraphics[width=.38\textwidth,trim={0 5pt 0 0}]{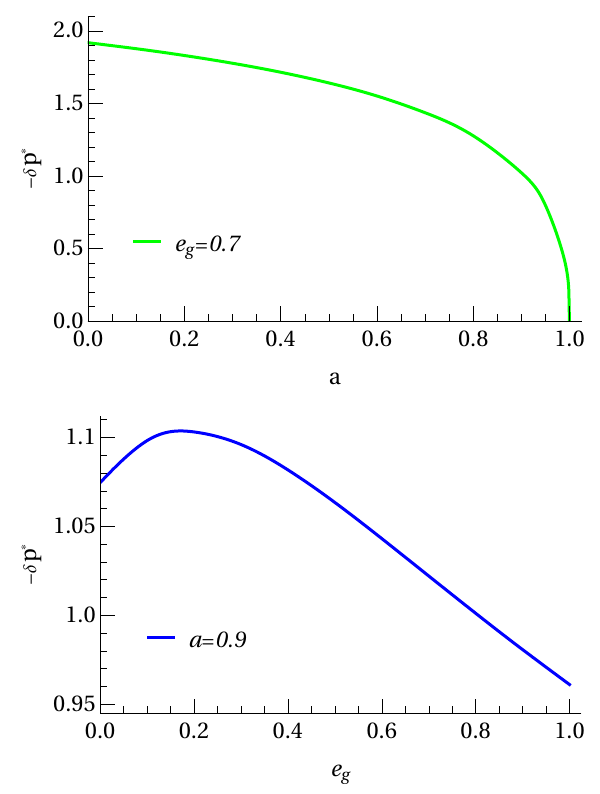}}
        \caption{Spin corrections to the location of the separatrix $\delta p^*$ for different orbital parameters. Top panel: $\delta p^*$ as a function of the primary spins $a$ for fixed geodesic eccentricity $e_{\rm g} =0.7$. Bottom panel: $\delta p^*$ for different values of geodesic eccentricities $e_{\rm g}$ and primary spin $a=0.9$. 
}       \label{fig:correction_separatrix}
\end{figure}

From now on, we only consider the FE gauge. For $r_{2 \rm g} = r_{3 \rm g}$ (i.e. at the geodesic separatrix), Eq.~\eqref{eq:semi_latus_rectum_shift} represents \textit{the spin correction to the separatrix location}, which we denote with $\delta p^*$. Such quantity depends only on the primary spin $a$ and the eccentricity at the geodesic separatrix $e^*_{\rm g}$. Fig.~\ref{fig:correction_separatrix} presents plots of $\delta p^*(a,e^*_{\rm g})$ for fixed $e_{\rm g}$ (top panel) and fixed $a$ (bottom panel). In Schwarzschild spacetime, $\delta p^*$ reduces to
\begin{equation}
    \delta p^*(0, e^*_{\rm g}) = - 2 \sqrt{\frac{2(1 + e^*_{\rm g})}{3 + e^*_{\rm g}}}
\end{equation}
which agrees with Ref.~\cite{Witzany:2023bmq}. Moreover, the shift to the ISCO is given by $\delta p^*_\text{ISCO} \equiv \delta p^*(a,0)$, which agrees with Jefremov et al.~\cite{Jefremov:2015gza}. Interestingly, the spin correction to the separatrix location vanishes for an extremal SMBH, whereas it is finite for $e^*_{\rm g} =1$ (which represents a marginally bound orbit).

\begin{figure*}[!bth]
    {\centering
        \includegraphics[width=\textwidth]{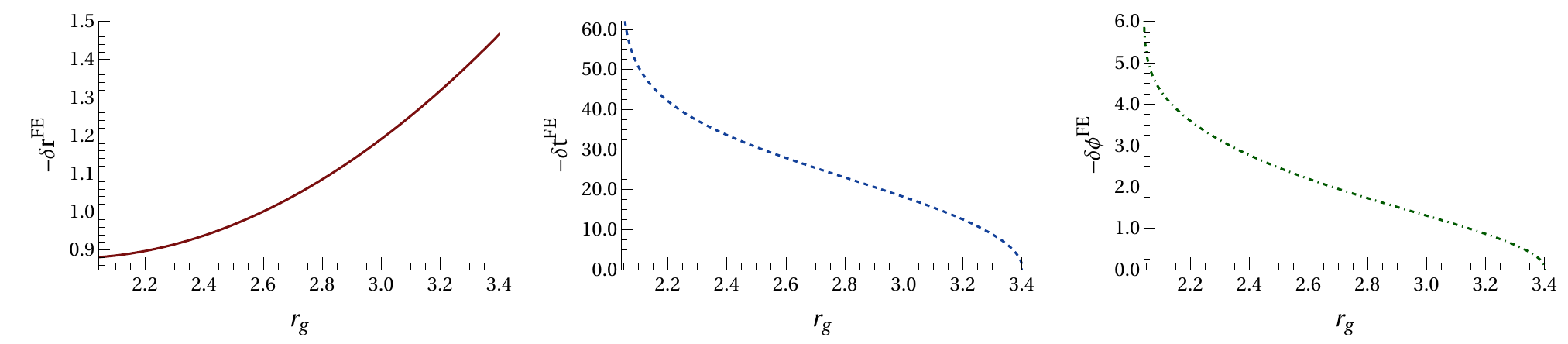}}
        \caption{Spin corrections to radial (left panel), coordinate time (middle panel) and azimuthal trajectories (right panel) for homoclinic, prograde orbits for a spinning binary with $a =0.9$, $\chi_\parallel = 1/2$, $\chi_\perp =\sqrt{3}/2$. All shifts to the trajectories are computed in the FE gauge. The underlying fiducial geodesic has eccentricity $e_{\rm g} =0.25$.
}       \label{fig:shift_homoclinic_orbit}
\end{figure*}

Aided by the limits of Appendix~\ref{app:geo_homoclinic} and the integrals of Appendix~\ref{app:integrals_periodic_motion}, the homoclinic trajectories can be obtained by simply taking the limit $r_{3 \rm g} \to r_{2 \rm g}$ of Eqs.~\eqref{eq:radial_shift_trajectory_periodic}-\eqref{eq:azimuthal_shift_trajectory_periodic}, which become
\begin{align} 
        \delta r^\FE(r_{\rm g}) &= \sqrt{R_g(r_{\rm g})}\bigg[ \bigg(\frac{E_{\rm g}\delta E^\FE}{1 - E^2_{\rm g}} + \frac{\delta \Upsilon^\FE_r}{\Upsilon_{r\rm g}}\bigg) I^\hom(r_{\rm g})  \nonumber \\
        & + \displaystyle \sum^2_{i=1} I^{\hom}_{1/(r - r_i)}(r_{\rm g})\frac{\delta r^\FE_i}{2} - \frac{a}{2} I^\hom_{1/r^2}(r_{\rm g})  \nonumber \\
        & + \delta r_{4} I^\hom_{1/r}(r_{\rm g})  \bigg]\ , \label{eq:rad_shift_trajectory_homoclinic}
\end{align}
\begin{widetext}
    \begin{align}
        \delta t^\FE(r_{\rm g}) &= - \frac{T_{\rm g}(r_{\rm g})\delta r^\FE(r_{\rm g})}{\sqrt{R_{\rm g}(r_{\rm g})}} + \displaystyle \sum^2_{i=1} T_{\rm g}(r_{i \rm g}) \frac{\delta r^\FE_i}{2} I^{\hom}_{1/(r - r_i)}(r_{\rm g}) + \frac{\delta E^\FE}{1 - E^2_{\rm g}} \Big(I^\hom_{r^2}(r_{\rm g}) + 2 I^\hom_{r}(r_{\rm g}) + 4 I^\hom(r_{\rm g}) \Big) \nonumber \\
       &  + \frac{E_{\rm g}}{2} \Big[ \big( \delta r^\FE_1 + 2 \delta r^\FE_2 + 2 \delta r_4 \big) I^\hom_r(r_{\rm g}) + \big((2 + r_{1 \rm g})\delta r^\FE_1 + 2(2 + r_{2 \rm g})\delta r^\FE_2 - a + 4\delta r_4 \big) I^\hom(r_{\rm g}) \Big]   \ , \label{eq:time_shift_trajectory_homoclinic}
    \end{align}
\end{widetext}    
    \begin{align}
        \delta \phi^\FE(r_{\rm g}) &= - \frac{\Phi_{\rm g}(r_{\rm g})\delta r^\FE(r_{\rm g})}{\sqrt{R_{\rm g}(r_{\rm g})}}  \nonumber \\
        & - \bigg(E_{\rm g} - L_{z \rm g}\frac{E_{\rm g} \delta E}{1 - E^2_{\rm g}} - \delta L_z \bigg) I^\hom(r_{\rm g})  \nonumber \\
        &+ \displaystyle \sum^2_{i=1} \Phi_{\rm g}(r_{i \rm g}) \frac{\delta r^\FE_i}{2} I^{\hom}_{1/(r - r_i)}(r_{\rm g})    \ , \label{eq:azimuthal_shift_trajectory_homoclinic}
    \end{align}    
Appendix~\ref{app:integrals_homoclinic_orbits} lists the integrals $I^\hom(r_{\rm g})$, $I^\hom_{1/r}(r_{\rm g})$, $I^\hom_{1/r^2}(r_{\rm g})$, $I^\hom_{r}(r_{\rm g})$, $I^\hom_{r^2}(r_{\rm g})$, which are combinations of elementary functions.  Fig.~\ref{fig:shift_homoclinic_orbit} presents the shifts to the homoclinic trajectories due to the secondary spin. Notice that $\delta t^\FE(r_{\rm g})$ and $\delta \phi^\FE(r_{\rm g})$ have the same logarithmic divergence at $r_{\rm g} = r_{2 \rm g}$ of their geodesic counterparts.

\begin{figure}[!bth]
    {\centering
        \includegraphics[width=0.43\textwidth]{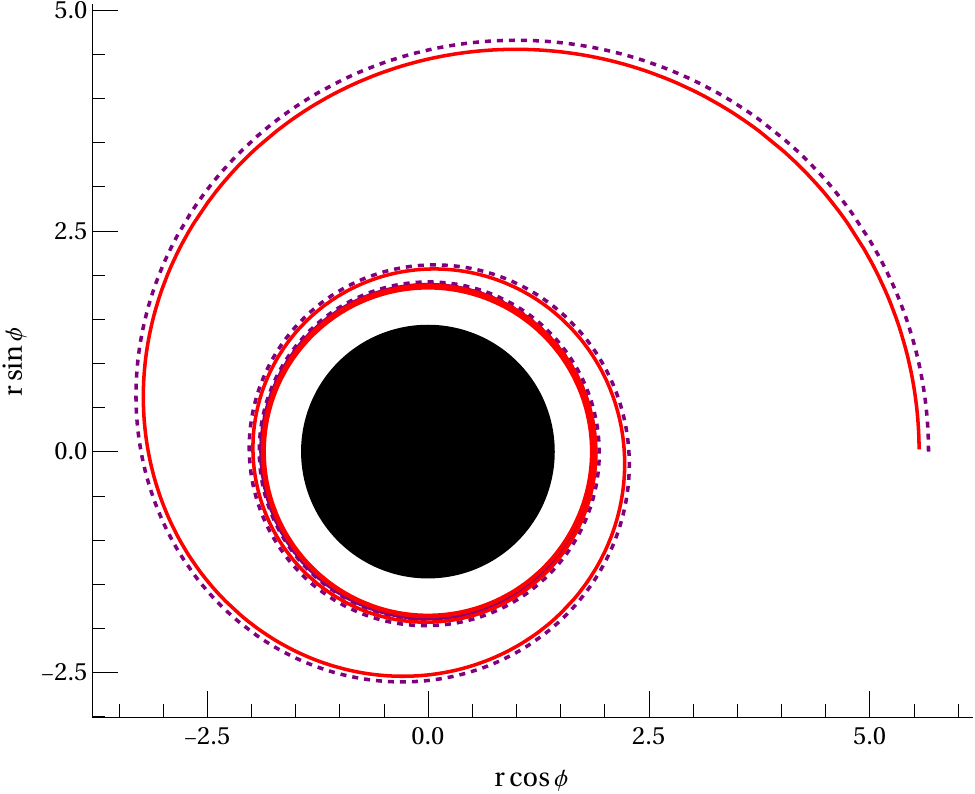} \\[0.4em]
        \includegraphics[width=0.43\textwidth]{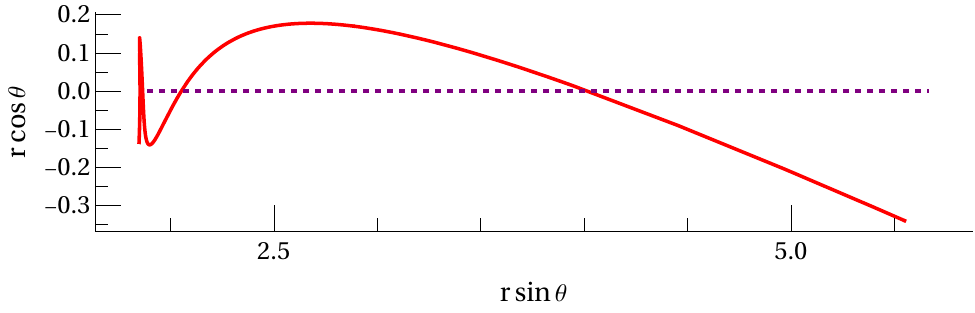}
    }
        \caption{Homoclinic orbits for a spinning particle (red-line) and its underlying fiducial geodesic (purple, dashed line).  The orbital parameters and the linear-in-spin corrections are the same of Fig.~\eqref{fig:shift_homoclinic_orbit}, while the mass-ratio is fixed to $q = 1/10$. The radius of the black disc corresponds to the SMBH outer horizon $r_+ = 1+\sqrt{1 -a^2}$. Top panel:  projection of the orbits onto the equatorial plane. Bottom panel: orthogonal projection of the orbits onto a co-rotating polar plane.
}       \label{fig:homoclinic_orbit}
\end{figure}

Finally, by taking the limit $e^*_{\rm g} \to 0$, the shift to the turning points collapse to $\delta r^\FE_1 = \delta r^\FE_2 = \delta p^*_\text{ISCO}$, while Eq.~\eqref{eq:rad_shift_trajectory_homoclinic} continuously reduces to $\delta p^*_\text{ISCO}$.

\section{Polar motion and spin-precession} \label{sec:polar_motion_spin_prec}
The polar motion is governed by Eq.~\eqref{eq:polar_EoM}, which describes the forced harmonic oscillations driven by the spin-curvature coupling. General solutions of Eq.~\eqref{eq:polar_EoM} can be written as the superposition of an homogeneous solution $\delta z_{\text{h}}$ and a particular solution $\delta z_{\text{inh}}$, given by
    \begin{subequations}
	\begin{empheq}[]{align}\label{eq:polar_sol}
		\delta z(\lambda) & = \delta z_{\text{h}}(\lambda) + \delta z_{\text{inh}}(\lambda) \ ,  \\
		\delta z_{\text{h}}(\lambda) &=  A \cos(\Upsilon_z \lambda) + B \sin(\Upsilon_z \lambda) \ , \\
        \delta z_{\text{inh}}(\lambda) &= - \chi_\perp \frac{\cos\psi_{\rm p} \sqrt{r^2_{\rm g} + (L_{z\rm g} - a E_{\rm g})^2}}{r_{\rm g} (L_{z\rm g} - a E_{\rm g})}  \ ,  \label{eq:polar_particular_sol} 
	\end{empheq}
	\end{subequations}
where $A$ and $B$ are constants fixed by the initial conditions. The referential geodesics is constrained on the equatorial plane when $A=B=0$, which can be imposed with the initial conditions $(\delta z_\text{h}(0),\dd \delta z_\text{h}(0)/ \dd \lambda) = (0,0)$. The particular solution $\delta z_\text{inh}(\lambda)$ was found in~\cite{Witzany:2023bmq} for a non-rotating primary, and it was later extended to Kerr spacetime in~\cite{Piovano:2024yks}. Moreover, Eq.~\eqref{eq:polar_particular_sol} agrees with the results of~\cite{Skoupy:2024uan}.

A few comments are in order. First, Eq.~\eqref{eq:polar_particular_sol} describes the precession of the orbital plane driven by the component of the particle spin orthogonal to the SMBH spin. Noteworthy, the complete solution~\eqref{eq:polar_sol} is valid for any type of radial trajectory $r_{\rm g}$ and precession phase $\psi_p$ constrained on the equatorial plane. Additionally, the general solution~\eqref{eq:polar_sol} do not admit resonant motion for any physical orbits. Such behavior was observed by Drummond and Hughes in~\cite{Drummond:2022_near_eq} for circular and nearly circular orbits. Moreover, the shift $\delta z(\lambda)$ does not depend on the corrections to the constants of motion. Thus, the orbital precession is the same regardless of the chosen referential geodesic.

A complete description of the polar motion requires the spin precession phase $\psi_p(\lambda)$, which evolves according to Eq.~\eqref{eq:spin-precession-angle}. The evolution equation for $\psi_p(r_{\rm g})$ can be recasted as
\begin{equation}
    \totder{\psi_p}{r_{\rm g}} = \pm \frac{\Psi_r(r_{\rm g})}{\sqrt{R_{\rm g}(r_{\rm g})}}  \ , 
\end{equation}
A periodic solution of the former equation is
\begin{align}
    \psi^\per_p(r_{\rm g}) &= \big(a\sgn(\Lzrd) + E_{\rm g}|\Lzrd|\big) I^\per(r_{\rm g})  \nonumber \\ 
    &- \frac{\Lzrd}{2} (a + E_{\rm g}\Lzrd) \Im \!\Big( I^\per_{1/(r - i |\Lzrd|)}(r_{\rm g}) \nonumber \\
    & - I^\per_{1/(r + i |\Lzrd|)}(r_{\rm g}) \Big) \ , \label{eq:spin_precession_phase_periodic}
\end{align}
where the integrals $I^\per(r_{\rm g}), I^\per_{1/(r \pm i |\Lzrd|)}(r_{\rm g})$ are given in Appendix~\ref{app:integrals_periodic_motion}, while the spin precession frequency is
\begin{equation}
    \Upsilon_p = \frac{\Upsilon_{r\rm g}}{\pi} \psi_p(r_{2\rm g}) \ ,    
\end{equation}
The precession phase $\psi_p(r_{\rm g})$ for homoclinic motion can be obtained by taking the limit $r_{3 \rm g} \to r_{2 \rm g}$ in Eq.~\eqref{eq:spin_precession_phase_periodic}, which reduces to
\begin{align}
    \psi^\hom_p(r_{\rm g}) &= \big(a\sgn(\Lzrd) + E_{\rm g}|\Lzrd|\big) I^\hom(r_{\rm g})  \nonumber \\ 
    &- \frac{\Lzrd}{2} (a + E_{\rm g}\Lzrd) \Im \!\Big( I^\hom_{1/(r - i |\Lzrd|)}(r_{\rm g}) \nonumber \\
    & - I^\hom_{1/(r + i |\Lzrd|)}(r_{\rm g}) \Big) \ ,  \label{eq:spin_precession_phase_homoclinic}
\end{align}
with the integrals $I^\hom(r_{\rm g}), I^\hom_{1/(r \pm i |\Lzrd|)}(r_{\rm g})$ given in Appendix~\ref{app:integrals_homoclinic_orbits}.
Moreover, the spin precession frequency at the separatrix reduces to $\Upsilon_p = \Psi_r(r_{2\rm g})$.

The bottom panels of Fig.~\ref{fig:zoom_whirl_orbit} and Fig.~\ref{fig:homoclinic_orbit} present the orthogonal projection onto a co-rotating polar plane for periodic and homoclinic orbits, respectively. Notice the precession of the orbital plane for the spinning trajectory (red line) due to the orthogonal component of the spin $\chi_\perp$.

\section{Discussion}\label{sec:discussion}
In this work, we studied the nearly-equatorial motion of a spinning test-particle in Kerr spacetime at linear order in the small spin. We derived analytic solutions for periodic orbits valid in any spin gauge. Our findings are in perfect agreement with the numerical trajectories of Refs.~\cite{Drummond:2022_near_eq,Piovano:2024yks}, which were derived in the fixed turning points and fixed constants of motion parametrizations. Furthermore, we introduced the novel fixed eccentricity parametrization, which is the unique spin gauge in which the corrections to periodic orbits are finite at the geodesic separatrix. The FE gauge allowed us to calculate, for the first time, the spin corrections to homoclinic orbits and the shift to the separatrix location in closed form.

A key difference from Ref.~\cite{Skoupy:2024uan} is that our solutions are given in terms of physical geodesic plus spin corrections in Mino-time. Thus, the results of this work can be readily applied to modeling the inspiral~\cite{Piovano:2024yks,Skoupy:2023lih} and transition to plunge~\cite{Becker:2024xdi,Kuchler:2024esj,Lhost:2024jmw,Compere:2021zfj,Compere:2021iwh} for spinning EMRI binaries, or to derive Post-Newtonian (PN) expressions for the dynamics~\cite{Henry:2023tka,Skoupy:2024jsi}. 

The analytic solutions for periodic orbits can be used to compute the spin-corrections to the gravitational fluxes and waveform amplitudes more efficiently than numerical schemes~\cite{Skoupy:2023lih}. Numerical integration methods typically struggles in the presence of singularities, such as the double root at the separatrix. Our homoclinic solutions are given as elementary functions, therefore they can be computed accurately and cheaply, making them easy to incorporate into SF frameworks for the calculation of transition-to-plunge waveforms~\cite{Kuchler:2024esj,Honet:2025dho}. Being in closed form, the periodic solutions can be also easily expanded near the separatrix to derive spin corrections to the late-time inspiral~\cite{Lhost:2024jmw}.

In the transition-to-plunge dynamics, both linear-in-spin and conservative, first-order in the mass ratio SF effects enter at third post-leading transition-to-plunge (or PLT) order. It is necessary to expand the field equations up to 7 PLT order to obtain complete inspiral-merger-ringdown (IMR) waveforms via SF schemes~\cite{Kuchler:2024esj,Honet:2025dho}. This approach would require the knowledge of second order in the mass-ratio SF and $\mathcal O(q^2 \chi^2)$ spin terms. Thus, our linear-in-spin homoclinic solutions provide an important ingredient to IMR-SF waveforms , but quadratic-in-spin corrections to homoclinic motion are needed to construct complete models.

Our approach relies on the partial separability of the equations of motion, which only occurs for near equatorial orbits, and quasi-spherical, periodic motion~\cite{Skoupy:2025nie}. However, the solutions given by Eqs.~\eqref{eq:generic_solution_radial_shift}-\eqref{eq:polar_particular_sol} and Eq.~\eqref{eq:polar_sol} are also valid for nearly equatorial scattering and plunging orbits. In a follow-up work~\cite{PiovanoInPrep}, we will derive explicit spin corrections to nearly equatorial bound, plunging orbits, in a similar vein to the geodesic case~\cite{Dyson:2023fws,Ciou:2025ygb,Mummery:2023hlo,Ko:2023igf}. Closed-form solutions for spinning trajectories would find practical use in the modeling of comparable mass binaries through BH perturbation techniques~\cite{Kuchler:2025hwx,Faggioli:2025hff,Compere:2021zfj,Compere:2021iwh}. Our linear-in-spin analytical solutions are sufficiently accurate for waveform models known at 1PA order. As shown in Refs.~\cite {Wardell:2021fyy,Mathews:2025nyb}, complete 1PA waveform models are in remarkable agreement with numerical relativity (NR) simulations throughout most of the inspiral phase. Hybrid PN-SF waveform models achieve even smaller mismatches with NR simulations, almost up to merger~\cite{Honet:2025lmk,Honet:2025gge}. These hybrid PN-SF models incorporate higher-order post-adiabatic terms obtained from PN expansions, which include $\mathcal O(q^2 \chi^2)$ and higher-order secondary spin effects. Thus, our analytical solutions are not sufficiently accurate for SF waveforms targeting comparable mass-binaries. For these binaries, it would be necessary to solve the MPD equations up to $\mathcal O(q^2 \chi^2)$ order, including spin-induced quadrupole effects. 

Our solutions could be useful to study test particles equipped with a spin-induced quadrupole on eccentric orbits, since they can used as input for perturbative solutions of the MPD equations. First-order equations of motion for a quadrupolar test-body are already available in the literature, at least on the equatorial plane~\cite{Bini:2013uwa,Bini:2013rrx,Bini:2015zya}. However, a careful study is necessary, as the dynamics at quadratic order in the spin is more involved~\cite{Steinhoff:2009tk,Steinhoff:2012rw,Shahzadi:2025ebj}. For example, the four-linear momentum and velocity are no longer colinear at order $\mathcal O(q^2 \chi^2)$. 

It is also possible to find analytic solutions for quasi-spherical periodic motion without using the transformations of~\cite{Skoupy:2024uan}, since the equations of motion in this case are still partially decoupled. The FE gauge can also be extended to quasi-spherical and generic orbits by introducing appropriate averaging~\cite{Piovano:2024yks}.

Finally, it would be interesting to find the transformation between the numerical solutions found in~\cite{Skoupy:2022adh,Drummond:2022efc,Piovano:2024yks}, and the analytic expressions given in the ``virtual geodesic" formalism of~\cite{Skoupy:2024uan}. The solutions found here would be the perfect starting point for such a comparison.

\begin{acknowledgments}
This work makes use of the following \textit{Mathematica} packages of the Black Hole Perturbation Toolkit~\cite{BHPToolkit}: ``KerrGeodesic"\cite{BHPToolkitKerrGeodesics}.
All supplementary material are available in the GitHub repository~\cite{repoAnalyticHJproject}.
G.A.P. is grateful to Josh Mathews, Lisa Drummond, Scott Hughes, Viktor Skoup\'y  and Vojt\u ech Witzany for many insightful discussions.
G.A.P. also acknowledges the support of the Win4Project grant ETLOG of the Walloon Region for the Einstein Telescope and of the Institut Complexys of the University of Mons.

\end{acknowledgments}

\newpage


\appendix

\section{Integrals for periodic motion} \label{app:integrals_periodic_motion}
Table~\ref{tab:elliptic_integrals} lists our conventions for the Legendre elliptic integrals and Jacobi elliptic functions, which follow the Mathematica conventions.
\begin{table}[h!]
  \begin{center}
    \caption{Elliptic Integral Conventions}
    \label{tab:elliptic_integrals}
    \begin{tabular}{l|c}
      \toprule 
      \textbf{Function} & \textbf{Definition} \\
      \midrule 
       $\mathsf{F}(x|m)$& $\int_{0}^{x}\frac{1}{\sqrt{1-m\sin^{2}(\theta)}}\mathrm{d}\theta$\\
       $\mathsf{K}(m)$& $\mathsf{F}(\pi/2|m)$\\
       $\mathsf{E}(x|m)$& $\int_{0}^{x}\sqrt{1-m\sin^{2}(\theta)}\mathrm{d}\theta$\\
       $\mathsf{E}(m)$& $\mathsf{E}(\pi/2|m)$\\
        $\mathsf{\Pi}(n,x|m)$&$\int_{0}^{x}\frac{1}{(1-n\sin^{2}(\theta))\sqrt{1-m\sin^{2}(\theta)}}\mathrm{d}\theta$\\
        $\mathsf{\Pi}(n|m)$&$\mathsf{\Pi}(n,\pi/2|m)$\\
       $\mathsf{am}(u|m)$& $u=\mathsf{F}(\mathsf{am}(u|m)|m)$\\
       $\mathsf{sn}(u|m)$& $\sin(\mathsf{am}(u|m))$\\
      \bottomrule 
    \end{tabular}
  \end{center}
\end{table}
The $m$ argument in all the elliptic integrals and functions is $k_{r\rm g}$
\begin{align}
    k_{r \rm g} &= \frac{(r_{1\rm g} - r_{2\rm g})r_{3\rm g}}{(r_{1\rm g} - r_{3\rm g})r_{2\rm g}} \, , 
\end{align}
while the elliptic characteristic $n$ in $\mathsf{\Pi}(n|m)$ is one of the following terms
\begin{align}
    \gamma_r &= 1- \frac{r_{1\rm g}}{r_{2\rm g}}   \,,\\
    h_\alpha &= \frac{(r_{1 \rm g} - r_{2 \rm g})\alpha}{(r_{1 \rm g} - \alpha)r_{2\rm g}} \, ,
\end{align}
with $\alpha \in \mathbb{C}$. Moreover, the elliptic amplitude $x$ of Table~\eqref{tab:elliptic_integrals} is given by
\begin{equation}
    \xi(r_{\rm g}) = \arcsin{\!\Bigg(\sqrt{\frac{(r_{1 \rm g} - r_{\rm g})r_{2 \rm g}}{(r_{1 \rm g} - r_{2 \rm g})r_{\rm g}}}\Bigg)} \ . \label{eq:elliptic_amplitude_arcsin}
\end{equation}
The standard domain for the elliptic amplitude $x$ is $[0,\pi/2]$. However, this domain can be analytical extended to $[ -\pi/2 , \pi/2 ]$  by using the fact that all elliptic integrals are antisymmetric in the $x$ argument: $ \mathsf{F}(-x|m) = -\mathsf{F}(x|m)$, and so on. In our case, $0 \leq \xi \leq  \pi/2$ when the radial velocity is positive, while $-\pi/2 \leq \xi \leq  0$ when the radial velocity is negative. Since $ k_{r \rm g} \in [0,1]$, the elliptic amplitude $\xi$ can also be expressed using the Jacobi amplitude $\mathsf{am(u|m)}$ as
\begin{equation}
    \xi(\lambda) = \mathsf{am}\Big(\frac{1}{2}Y_r \lambda \Big \rvert k_{r \rm g}\Big) \ ,  \label{eq:elliptic_amplitude_am}
\end{equation}
where $Y_r = \sqrt{(1 - E^2_{\rm g})(r_{1 \rm g} - r_{3 \rm g})r_{2 \rm g}}$. Unlike Eq.~\eqref{eq:elliptic_amplitude_arcsin}, the previous expression for $\xi$ is finite for $r_{2 \rm g} = r_{1 \rm g}$, since $\mathsf{am}(u|0) = u$. 

We list here the integrals that appear in the analytic solutions for periodic orbits:
\begin{align}
    I^\per(r_{\rm g}) = \int^{r_{1\rm g}}_{r_{\rm g}} \frac{\dd r'}{\sqrt{R_{\rm g}(r')}} &= \frac{2\mathsf{F}(\xi| k_{r \rm g})}{Y_r} = \lambda  \ , \\
    I^\per_{r}(r_{\rm g}) = \int^{r_{1\rm g}}_{r_{\rm g}} \frac{r'\dd r'}{\sqrt{R_{\rm g}(r')}} &= \frac{2 r_{1 \rm g}}{Y_r}   \mathsf{\Pi}\big(\!\left.\gamma_r,\xi\right\rvert k_{r \rm g}\!\big) \ , 
\end{align}
\begin{widetext}
\begin{align}
    I^\per_{1/r}(r_{\rm g}) = \int^{r_{1\rm g}}_{r_{\rm g}} \frac{\dd r'}{r'\sqrt{R_{\rm g}(r')}} &= \frac{I^\per(r_{\rm g})}{r_{3 \rm g}} - \frac{2 (r_{1 \rm g} - r_{3 \rm g})\mathsf{E}(\xi| k_{r \rm g})}{r_{1 \rm g} r_{3 \rm g}Y_r}  \ , \\
    I^\per_{1/(r - r_3)}(r_{\rm g}) = \int^{r_{1\rm g}}_{r_{\rm g}} \frac{\dd r'}{(r' - r_{3 \rm g})\sqrt{R_{\rm g}(r')}} &= - \frac{I^\per(r_{\rm g})}{r_{3 \rm g}} + \frac{2 r_{2 \rm g}\mathsf{E}(\xi| k_{r \rm g})}{(r_{2 \rm g} - r_{3 \rm g})r_{3 \rm g}Y_r} - \frac{2\cos \xi \sin \xi}{(r_{2 \rm g} - r_{3 \rm g})Y_r\sqrt{1 - k_{r \rm g}\sin^2(\xi)}}  \ , \\
    I^\per_{1/r^2}(r_{\rm g}) = \int^{r_{1\rm g}}_{r_{\rm g}} \frac{\dd r'}{r'^2\sqrt{R_{\rm g}(r')}} &= - \frac{I^\per(r_{\rm g})}{r^2_{1 \rm g}} + \frac{2 I^\per_{1/r}(r_{\rm g})}{r_{1 \rm g}} + \frac{\gamma^2_r}{3 k^2_{\rm g} r^2_{1 \rm g}}(2 + k_{r \rm g})I^\per(r_{\rm g}) - \gamma^2_r\frac{2(1 + k_{r \rm g})\mathsf{E}(\xi| k_{r \rm g})}{3 k^2_{\rm g} r^2_{1 \rm g}Y_r}   \nonumber \\
    & + \frac{2}{3}\gamma^2_r\frac{\cos\xi \sin\xi \sqrt{1 - k_{r \rm g}\sin^2(\xi)}}{k_{r \rm g} r^2_{1 \rm g}Y_r}  \ , \\
    I^\per_{r^2}(r_{\rm g}) = \int^{r_{1\rm g}}_{r_{\rm g}} \frac{r'^2\dd r'}{\sqrt{R_{\rm g}(r')}} &=  - \frac{r_{1 \rm g} r_{2 \rm g}}{2}I^\per(r_{\rm g}) + \frac{Y_r}{(1 - E^2_{\rm g})} \mathsf{E}(\xi| k_{r \rm g}) + \frac{2r_{1 \rm g}\mathsf{\Pi}\big(\!\left.\gamma_r,\xi\right\rvert k_{r \rm g}\!\big) }{(1 - E^2_{\rm g})^2 Y_r}  \nonumber \\
    & + \frac{ (r_{1 \rm g} - r_{2 \rm g})Y_r \cos\xi \sin\xi \sqrt{1 - k_{r \rm g}\sin^2(\xi)}}{(1 - E^2_{\rm g}) (r_{2 \rm g} + (r_{1 \rm g} - r_{2 \rm g}))\sin^2(\xi)} \ , \\
    I^\per_{1/(r - \alpha)}(r_{\rm g}) &= \int^{r_{1\rm g}}_{r_{\rm g}} \frac{\dd r'}{(r' - \alpha)\sqrt{R_{\rm g}(r')}} = \frac{2 r_{1 \rm g} \mathsf{\Pi}\big(\!\left. h_\alpha,\xi\right\rvert k_{r \rm g}\!\big)}{\alpha (r_{1 \rm g} - \alpha)Y_r} - \frac{I^\per(r_{\rm g})}{\alpha} \ ,  
\end{align} 
\end{widetext}
Additionally, we used the following diverging integrals
\begin{align}
    I^\per_{1/(r - r_1)}(r_{\rm g}) &= \int^{r_{1\rm g}}_{r_{\rm g}} \frac{\dd r'}{(r' - r_{1 \rm g})\sqrt{R_{\rm g}(r')}} \nonumber \\
    & = I^{\per,\mathsf{sing}}_{1/(r - r_1)}(r_{\rm g}) + I^{\per,\mathsf{reg}}_{1/(r - r_1)}(r_{\rm g}) \ , \\
   I^\per_{1/(r - r_2)}(r_{\rm g}) &= \int^{r_{1\rm g}}_{r_{\rm g}} \frac{\dd r'}{(r' - r_{2 \rm g})\sqrt{R_{\rm g}(r')}} \nonumber \\
    & = I^{\per,\mathsf{sing}}_{1/(r - r_2)}(r_{\rm g}) + I^{\per,\mathsf{reg}}_{1/(r - r_2)}(r_{\rm g}) \ , 
\end{align}
which are separated into a diverging piece
\begin{align}
    I^{\per,\mathsf{sing}}_{1/(r - r_1)}(r_{\rm g}) &= \frac{2(r_{\rm g} - r_{2 \rm g})(r_{\rm g} - r_{3 \rm g})}{(r_{1 \rm g} - r_{2 \rm g})(r_{1 \rm g} - r_{3 \rm g})\sqrt{R_{\rm g}(r_{\rm g})}}   \ , \\
    I^{\per,\mathsf{sing}}_{1/(r - r_2)}(r_{\rm g}) &= \frac{2(r_{1\rm g} - r_{\rm g})(r_{\rm g} - r_{3 \rm g})}{(r_{1 \rm g} - r_{2 \rm g})(r_{2 \rm g} - r_{3 \rm g})\sqrt{R_{\rm g}(r_{\rm g})}} \ , 
\end{align}
and a regular part
\begin{align}
    I^{\per,\mathsf{reg}}_{1/(r - r_1)}(r_{\rm g}) &= \frac{2 r_{2 \rm g} \mathsf{E}(\xi|k_{r\rm g})}{r_{1 \rm g}(r_{1 \rm g} - r_{2 \rm g})Y_r} - \frac{I^\per(r_{\rm g})}{(r_{1 \rm g} - r_{2 \rm g})} \ , \\
    I^{\per,\mathsf{reg}}_{1/(r - r_2)}(r_{\rm g}) &= \frac{I^\per(r_{\rm g})}{(r_{1 \rm g} - r_{2 \rm g})} - \frac{2(r_{1 \rm g} - r_{3 \rm g}) \mathsf{E}(\xi|k_{r\rm g})}{(r_{1 \rm g} - r_{2 \rm g})(r_{2 \rm g} - r_{3 \rm g})Y_r}  \ .
\end{align}

\section{Integrals for homoclinic orbits} \label{app:integrals_homoclinic_orbits}
We list here the integrals that appear in the analytic solutions for homoclinic orbits: 
\begin{widetext}    
\begin{align}
    I^\hom(r_{\rm g}) &= \int^{r_{1 \rm g}}_{r_{\rm g}} \frac{\dd r'}{\sqrt{R_{\rm g}(r')}} = -\frac{1}{2\sqrt{(1 - E^2_{\rm g})(r_{1 \rm g} - r_{2 \rm g})r_{2 \rm g}}} \log \!\Bigg[\Bigg( \frac{\sqrt{r_{\rm g}(r_{1 \rm g} - r_{2 \rm g})} -\sqrt{r_{2\rm g}(r_{1 \rm g} - r_{\rm g})}}{\sqrt{r_{\rm g}(r_{1 \rm g} - r_{2 \rm g})} + \sqrt{r_{2 \rm g}(r_{1 \rm g} - r_{\rm g)}} } \Bigg)^{\!\!2} \Bigg] = \lambda\ , \\
    I^\hom_{r}(r_{\rm g}) &= \int^{r_{1 \rm g}}_{r_{\rm g}} \frac{r'\dd r'}{\sqrt{R_{\rm g}(r')}} = \frac{2}{\sqrt{1 - E^2_{\rm g}}} \arccos\bigg(\sqrt{\frac{r_{\rm g}}{r_{1\rm g}}} \bigg) + r_{2 \rm g} I^\hom(r_{\rm g}) \ , \\
    I^\hom_{1/r}(r_{\rm g})  & = \int^{r_{1 \rm g}}_{r_{\rm g}} \frac{\dd r'}{r'\sqrt{R_{\rm g}(r')}} =  - \frac{2\sqrt{r_{1 \rm g} - r_{\rm g}}}{r_{1 \rm g} r_{2 \rm g}\sqrt{(1 - E^2_{\rm g})r_{\rm g}}} + \frac{I^\hom(r_{\rm g})}{r_{2 \rm g}} \ , \\
    I^\hom_{1/r^2}(r_{\rm g}) &= \int^{r_{1 \rm g}}_{r_{\rm g}} \frac{\dd r'}{r'^2\sqrt{R_{\rm g}(r')}} = - \frac{2}{3} \frac{\sqrt{(r_{1 \rm g} - r_{\rm g})}}{r_{1 \rm g} r_{2 \rm g} \sqrt{(1 - E^2_{\rm g})r_{\rm g}}} \bigg[ \frac{2}{r_{1 \rm g}} + \frac{1}{r_{2 \rm g}} \bigg(3 + \frac{r_{2 \rm g}}{r_{\rm g}} \bigg) \bigg]  + \frac{I^\hom(r_{\rm g})}{r^2_{2\rm g}}  \ , \\
    I^\hom_{r^2}(r_{\rm g}) &= \int^{r_{1 \rm g}}_{r_{\rm g}} \frac{r'^2\dd r'}{\sqrt{R_{\rm g}(r')}} = \frac{\sqrt{(r_{1\rm g} - r_{\rm g})r_{\rm g}}}{\sqrt{1 - E^2_{\rm g}}} + \frac{r_{1\rm g} + 2 r_{2 \rm }}{\sqrt{1 - E^2_{\rm g}}}  \arccos\bigg(\sqrt{\frac{r_{\rm g}}{r_{1\rm g}}} \bigg) + r^2_{2 \rm g} I^\hom(r_{\rm g})   \ , \\
    I^\hom_{1/(r - \alpha)}(r_{\rm g}) &= \int^{r_{1 \rm g}}_{r_{\rm g}} \frac{\dd r'}{(r' - \alpha)\sqrt{R_{\rm g}(r')}} = \frac{I^\hom(r_{\rm g})}{r_{2 \rm g} - \alpha} + \frac{\log \!\Bigg[\Bigg( \frac{\sqrt{r_{\rm g}(r_{1 \rm g} -\alpha)} - \sqrt{\alpha(r_{1 \rm g} - r_{\rm g})}}{\sqrt{r_{\rm g}(r_{1 \rm g} - \alpha)} + \sqrt{\alpha(r_{1 \rm g} - r_{\rm g)}} } \Bigg)^{\!\!2} \Bigg]}{2(r_{2 \rm g} - \alpha)\sqrt{(1 - E^2_{\rm g})(r_{1 \rm g} - \alpha)\alpha}} 
\end{align}
\end{widetext}
Additionally, we used the following integrals
\begin{align}
    I^\hom_{1/(r - r_1)}(r_{\rm g}) &= \int^{r_{1\rm g}}_{r_{\rm g}} \frac{\dd r'}{(r' - r_{1 \rm g})\sqrt{R_{\rm g}(r')}} \nonumber \\
    & = I^{\hom,\mathsf{sing}}_{1/(r - r_1)}(r_{\rm g}) + I^{\hom,\mathsf{reg}}_{1/(r - r_1)}(r_{\rm g}) \ , \\
   I^\hom_{1/(r - r_2)}(r_{\rm g}) &= \int^{r_{1\rm g}}_{r_{\rm g}} \frac{\dd r'}{(r' - r_{2 \rm g})\sqrt{R_{\rm g}(r')}} \nonumber \\
    & = I^{\hom,\mathsf{sing}}_{1/(r - r_2)}(r_{\rm g}) + I^{\hom,\mathsf{reg}}_{1/(r - r_2)}(r_{\rm g}) \ , 
\end{align}
whose diverging part is given by
\begin{align}
    I^{\hom,\mathsf{sing}}_{1/(r - r_1)}(r_{\rm g}) &= \frac{2\sqrt{(r_{1 \rm g} - r_{\rm g})r_{\rm g}}}{r_{1 \rm g}(r_{1 \rm g} - r_{2 \rm g})(r_{1 \rm g} - r_{\rm g})}   \ , \\
    I^{\hom,\mathsf{sing}}_{1/(r - r_2)}(r_{\rm g}) &= \frac{\sqrt{(r_{1 \rm g} - r_{\rm g})r_{\rm g}}}{r_{2 \rm g}(r_{1 \rm g} - r_{2 \rm g})(r_{\rm g} - r_{2\rm g})} \ , 
\end{align}
while the regular term is defined as
\begin{align}
    I^{\hom,\mathsf{reg}}_{1/(r - r_1)}(r_{\rm g}) &= - \frac{I^\hom(r_{\rm g})}{(r_{1 \rm g} - r_{2 \rm g})} \ , \\
    I^{\hom,\mathsf{reg}}_{1/(r - r_2)}(r_{\rm g}) &=  - \frac{(r_{1 \rm g} - 2 r_{2 \rm g})I^\hom(r_{\rm g})}{2(r_{1 \rm g} - r_{2 \rm g})r_{2 \rm g}}   \ .
\end{align}
\section{Analytic solutions for geodesic motion on the equatorial plane} \label{app:geo_analytic}
For the reader's convenience, we present here the analytic solutions for geodesic equatorial motion in Mino-time with the initial conditions and conventions of Sec.~\ref{sec:geodesic_equatorial_motion}. Refs.~\cite{Fujita:2009bp,vandeMeent:2019cam} present the solutions for generic periodic geodesics using different initial conditions, while Ref.~\cite{Levin:2008yp} presents homoclinic geodesic in closed form parametrized using proper time (see Ref.~\cite{Li:2023bgn} for off-equatorial homoclinic in Kerr-Neuwmann spacetime). 

The starting points are the equations of motion~\eqref{eq:1st-order-EoMgeo}, which can be recasted in the following form
\begin{subequations}
   \begin{empheq}[]{align}
    &\lambda = \pm \int^{r_{1\rm g}}_{r_{\rm g}} \frac{\dd r'}{\sqrt{R_{\rm g}(r')}} \, ,  \\  
	&\totder{t}{r_{\rm g}} = \pm \frac{T_{\rm g}(r_{\rm g})}{\sqrt{R_{\rm g}(r_{\rm g})}} \, , \\ 
	&\totder{\phi}{r_{\rm g}} = \pm \frac{\Phi_{\rm g}(r_{\rm g})}{\sqrt{R_{\rm g}(r_{\rm g})}}  \, .
   \end{empheq}
\end{subequations} 
Analytic expressions for periodic and homoclinic geodesics are presented in the next subsections.
\subsection{Periodic trajectories}
For periodic orbits, the solution to the radial motion is given by
\begin{equation} 
    r_{\rm g}(\lambda) = \frac{r_{1 \rm g} r_{2 \rm g}}{r_{2 \rm g} + (r_{1 \rm g} - r_{2 \rm g}) \mathsf{sn}^2(\xi(\lambda)|k_{r \rm g})} \ , \label{eq:radial_geo_trajectory}
\end{equation}
which oscillates with Mino-time frequency $\Upsilon_{r\rm g}$
\begin{align}
	\Upsilon_{r\rm g} =\frac{\pi}{2\, \mathsf{K}(k_{r\rm g})}\sqrt{(1-E_{\rm g}^2)(r_{1\rm g}-r_{3\rm g})r_{2\rm g}}\,,
\end{align}
As shown in Ref.~\cite{vandeMeent:2020xgc}, the radial motion can then be parametrized using the geodesic mean anomaly $w_{r \rm g} = \Upsilon_{r\rm g} \lambda$ by taking advantage of the identity of the Jacobi amplitude $\mathsf{am}( \mathsf{K}(m)|m) = \pi/2$. Thus, the eccentric amplitudes $\xi$ can be written in terms of the mean anomaly as
\begin{equation}
    \xi(w_{r \rm g}) = \mathsf{am}\Big(\mathsf{K(k_{r \rm g})}\frac{w_{r \rm g}}{\pi} \Big \rvert k_{r \rm g}\Big) \ , \label{eq:elliptic_amplitude_of_wr}
\end{equation}
The geodesic radial potential can then be recasted using Eq.~\eqref{eq:radial_geo_trajectory} and Eq.~\eqref{eq:elliptic_amplitude_of_wr}  as
\begin{widetext}
    \begin{equation}
        \pm \sqrt{R_{\rm g}(r_{\rm g})} = \sqrt{1 - E^2_{\rm g}}\frac{(r_{1 \rm g} - r_{2 \rm g})(r_{1 \rm g} - r_{3 \rm g})(r_{2 \rm g} - r_{3 \rm g})\sin(\xi) \cos(\xi)\sqrt{r_{2 \rm g}(r_{1 \rm g} - r_{3 \rm g}) - (r_{1 \rm g} - r_{2 \rm g})r_{3 \rm g}\sin^2(\xi)}}{\big( (r_{1 \rm g} - r_{2 \rm g}) \sin^2(\xi) - r_{1 \rm g} + r_{3 \rm g}\big)^2 }  \ ,
    \end{equation}    
\end{widetext}
Finally, the coordinate time and azimuthal trajectories can be written with the aid of the integrals of Appendix~\ref{app:integrals_periodic_motion} as
\begin{widetext}
    \begin{align}
        t_{\rm g}(r_{\rm g}) &= E_{\rm g} \big(4 I^\per(r_{\rm g}) +  2 I^\per_{r}(r_{\rm g}) + I^\per_{r^2}(r_{\rm g}) \big) - \frac{2}{r_+ - r_-} \Big( 2 a^2 E_{\rm g} - 4 E_{\rm g} r_+ + a L_{z \rm g} r_+ \Big)I^\per_{1/(r - r_+)}(r_{\rm g}) + (+\leftrightarrow-)   \ ,  \label{eq:coordinate_time_geo_trajectory} \\ 
        \phi_{\rm g}(r_{\rm g}) &= L_{z\rm g} I^\per(r_{\rm g}) + \frac{a (2r_+ E_{\rm g} - a L_{z \rm g})}{(r_+ - r_-)} I^\per_{1/(r - r_+)}(r_{\rm g}) +  (+\leftrightarrow-)  \ , \label{eq:azimuthal_geo_trajectory}
    \end{align}    
\end{widetext}
while the corresponding geodesics frequencies $\Upsilon_{t \rm g}$ and $\Upsilon_{\phi \rm g}$ are given by
\begin{align}
    \Upsilon_{t \rm g} = \frac{\Upsilon_{r \rm g}}{\pi}t_{\rm g}(r_{2 \rm g}) \ , 
    \qquad \Upsilon_{\phi \rm g} = \frac{\Upsilon_{r \rm g}}{\pi}\phi_{\rm g}(r_{2 \rm g})  \ .
\end{align}
The purely oscillatory components of the geodesics $t_{\rm g}(r_{\rm g})$ and $\phi_{\rm g}(r_{\rm g})$ can then be expressed as
\begin{align}
    b_\text{\rm g,osc}(w_{r\rm g})  &= b_{\rm g}(w_{r\rm g}) - \frac{\Upsilon_{b \rm g}}{\Upsilon_{r\rm g}} w_{r_{\rm g}} \ ,
\end{align}
with $b = t,\phi$.
    
\subsection{Homoclinic trajectories} \label{app:geo_homoclinic}
Homoclinic trajectories can be obtained by taking the limit $r_{3 \rm g} \to r_{2 \rm g}$ of periodic geodesics. In this case, the argument $k_{r \rm g} = 1$ in the integrals in Appendix~\ref{app:integrals_periodic_motion}, which leads to 
\begin{align}
    \sin(\xi) &= \mathsf{sn}\Big(\frac{1}{2}Y_r \lambda \Big \rvert 1\Big) = \tanh{\Big(\frac{1}{2}Y_r \lambda \Big)}  \ , \\
    \mathsf{F}(\xi|1) &= \log\big(\sec(\xi) + \tan(\xi)\big) = \frac{1}{2}Y_r \lambda \ , \\
    \mathsf{E}(\xi|1) &= \sin(\xi) \ , \\
    \mathsf{\Pi}(n,\xi|1) &= \frac{\sqrt{n}}{n-1}  \arctanh(\sqrt{n}\sin{\xi}) - \frac{Y_r \lambda }{2(n-1)} \ .
\end{align}
Using the previous expressions, the homoclinic radial trajectory can then be written as
\begin{equation} 
    r_{\rm g}(\lambda) = \frac{r_{1 \rm g} r_{2 \rm g}}{r_{2 \rm g} + (r_{1 \rm g} - r_{2 \rm g}) \tanh^2\big(\frac{1}{2}Y_r \lambda\big)} \ ,
\end{equation}
while the homoclinic coordinate time and azimuthal trajectories can be read from Eqs.~\eqref{eq:coordinate_time_geo_trajectory}-\eqref{eq:azimuthal_geo_trajectory} by replacing $I^\per(r_{\rm g})$ with $I^\hom(r_{\rm g})$, $I^\per_r(r_{\rm g})$ with $I^\hom_r(r_{\rm g})$ and so on.
As noticed in~\cite{Levin:2008yp}, the trajectories $t(r_{\rm g})$ and $\phi(r_{\rm g})$ have a logarithmic divergence near the periastron, since the particle take an infinite Mino (and proper) time to reach the double root $r_{2 \rm g} = r_{3 \rm g}$. Therefore, the radial frequency $\Upsilon_{r \rm g} \to 0$ as $r_{3 \rm g} \to r_{2 \rm g}$. By contrast, $\Upsilon_{t \rm g}$ and $\Upsilon_{\phi \rm g}$ are finite in the limit $r_{3 \rm g} \to r_{2 \rm g}$, and reduce to
\begin{align} 
    \Upsilon_{t \rm g} = T_{\rm g}(r_{2 \rm g}) \ , \qquad   \Upsilon_{t \rm g} = \Phi_{\rm g}(r_{2 \rm g}) \ ,  
\end{align}
which were derived using the following limits
\begin{align} \label{eq:limits_frequencies}
    & \mathsf{K}(k_{r \rm g}) \sim  - \frac{1}{2} \log(r_{2\rm g} - r_{3 \rm g}) \quad \text{ as } \quad r_{3 \rm g} \to r_{2 \rm g}  \ ,\\
    & \lim_{r_{3 \rm g} \to r_{2 \rm g}} \frac{\Pi(\gamma_r|k_{r \rm g})}{\mathsf{K}(k_{r \rm g})} = \frac{1}{1 - \gamma_r} \ .
\end{align}

\section{Corrections to the constants of motion in the FT gauge} \label{app:spin_constants_of_motion_corr}
The spin corrections to the energy and angular momentum in the fixed turning points gauge are given by
\begin{align}
    \delta E^\FT & = \frac{R_s(r_{2\rm g})}{\mathcal J}\frac{\partial R_{\rm g}}{\partial L_{z \rm g}}(r_{1 \rm g}) - \frac{R_s(r_{1\rm g})}{\mathcal J}\frac{\partial R_{\rm g}}{\partial L_{z \rm g}}(r_{2 \rm g})  \ , \\
    \delta L^\FT_z & = \frac{R_s(r_{1\rm g})}{\mathcal J}\frac{\partial R_{\rm g}}{\partial E_{\rm g}}(r_{2 \rm g}) - \frac{R_s(r_{2\rm g})}{\mathcal J}\frac{\partial R_{\rm g}}{\partial E_{\rm g}}(r_{1 \rm g}) \ ,
\end{align}
with 
\begin{equation}
    \mathcal J = \frac{\partial R_{\rm g}}{\partial E_{\rm g}}(r_{1 \rm g}) \frac{\partial R_{\rm g}}{\partial L_{z\rm g}}(r_{2 \rm g}) - \frac{\partial R_{\rm g}}{\partial L_{z\rm g}}(r_{1 \rm g}) \frac{\partial R_{\rm g}}{\partial E_{\rm g}}(r_{2 \rm g}) \ .
\end{equation}
In these form, the shifts $\delta E^\FT$ and $\delta L_z^\FT$ are divergent for stable, circular geodesic, i.e. for $r_{1 \rm g} = r_{2 \rm g}$. However, after some algebraic effort, the shifts $\delta E^\FT$ and $\delta L_z^\FT$ can be written in the equivalent form
\begin{widetext}
    \begin{align}
        \delta E^\FT &=\frac{\Lzrd}{\mathcal D} \big[ -2 a \Lzrd^2(r_{1 \rm g} + r_{2 \rm g}) - L_{z\rm g} E_{\rm g} r^2_{1 \rm g} r^2_{2 \rm g} -a L_{z\rm g} \Lzrd r_{1 \rm g} r_{2 \rm g} + a \Lzrd L_{z\rm g} (r_{1 \rm g} + r_{2\rm g})^2 \big] \ , \label{eq:delta_E_FT} \\
        \delta L_z^\FT &=\frac{E_{\rm g}r_{1\rm g} r_{2\rm g}}{\mathcal D} \Lzrd \big[ a^3 \Lzrd - a (a E_{\rm g} - 3 \Lzrd) r_{1 \rm g} r_{2 \rm g} - 3 E_{\rm g} r^2_{1\rm g} r^2_{2\rm g} \big] \nonumber \\
        & + \frac{a \Lzrd^2}{\mathcal D}\big[ - 2 a\Lzrd (r_{1\rm g} + r_{2\rm g}) + a^2 E_{\rm g} (r^2_{1\rm g} + r^2_{2\rm g}) + E_{\rm g} (r^4_{1\rm g} + r^4_{2\rm g}) \big]  \nonumber\\            & + \frac{E_{\rm g}r_{1\rm g} r_{2\rm g}}{\mathcal D} \big[ E_{\rm g} L_{z \rm g}r^2_{1\rm g} r^2_{2\rm g}(r_{1\rm g} + r_{2\rm g}) - 3E_{\rm g}\Lzrd r_{1\rm g} r_{2\rm g}(r^2_{1\rm g} + r^2_{2\rm g}) + a \Lzrd^2 (r^2_{1\rm g} + r^2_{2\rm g}) \big]  \ ,  \label{eq:delta_Lz_FT}      
    \end{align}   
    with
    \begin{equation}
        \mathcal D = r^2_{1 \rm g} r^2_{2 \rm g} \Big[ E_{\rm g} L_{z\rm g} r_{1\rm g} r_{2\rm g} (r_{1\rm g} + r_{2\rm g}) + 2 a \Lzrd^2 - 2 \Lzrd  E_{\rm g}(r_{1\rm g} + r_{2\rm g})^2 + 2 \Lzrd  E_{\rm g}r_{1\rm g} r_{2\rm g}  \Big] \ .
    \end{equation}
\end{widetext}
Eqs.~\eqref{eq:delta_E_FT}-\eqref{eq:delta_Lz_FT} are well defined for stable and unstable circular orbits. In the former case, $\delta E^\FT$ and $\delta L_z^\FT$ reduced to the shifts of the constants of motion of~\cite{Jefremov:2015gza}.

\section{Corrections to the Mino-time frequencies for a spinning particle} \label{app:spin_frequency_corr}
This Section presents the analytic linear-in-spin corrections to the Mino-time frequencies. In a generic spin gauge, the shift to the radial motion $\delta \Upsilon_r$ can be defined as the finite part of Eq.~\eqref{eq:divergent_shift_radial_frequency} using Hadamard's partie finie~\cite{Witzany:2024ttz}. The singular part of the integral~\eqref{eq:divergent_shift_radial_frequency} is given by
\begin{widetext}
\begin{equation}
    \frac{\Upsilon^2_{r \rm g}}{\pi}\int^{r_{1 \rm g} - \epsilon}_{r_{2 \rm g} + \epsilon} \frac{\delta V_r(r_{\rm g}) \dd r'}{\sqrt{R_{\rm g}(r')}}  = \frac{\Upsilon^2_{r\rm g}}{\pi}\Big(I^{\per,\text{sing}}_{1/(r - r_1)}(r_{1 \rm g} - \epsilon)\delta r_1 + I^{\per,\text{sing}}_{1/(r - r_2)}(r_{2\rm g} + \epsilon)\delta r_2 \Big)
\end{equation}
for $\epsilon \to 0$, while the Hadamard's partie finie of Eq.~\eqref{eq:divergent_shift_radial_frequency} is
    \begin{align}
        \delta \Upsilon_r &= -\frac{\Upsilon^2_{r\rm g}}{\pi}\bigg[\frac{E_{\rm g}\delta E}{1 - E^2_{\rm g}} I^\per(r_{2 \rm g}) + \displaystyle \sum^2_{i=1} I^{\per,\text{reg}}_{1/(r - r_i)}(r_{2 \rm g})\frac{\delta r_i}{2} + I^\per_{1/(r - r_3)}(r_{2\rm g})\frac{\delta r_3}{2} - \frac{a}{2} I^\per_{1/r^2}(r_{2 \rm g}) + \delta r_{4} I^\per_{1/r}(r_{2 \rm g}) \bigg] \ , \label{eq:shift_radial_frequency}
    \end{align}
\end{widetext}
The shifts to Mino time frequencies $\delta \Upsilon_t$ and $\delta \Upsilon_\phi$ do not require regularization, and they can be written as
\begin{widetext}
    \begin{align}
        \delta \Upsilon_t &= \frac{\Upsilon_{t \rm g}}{\Upsilon_{r \rm g}} \delta \Upsilon_r + \frac{\Upsilon_{r \rm g}}{\pi}\bigg[ \displaystyle \sum^2_{i=1} T_{\rm g}(r_{i \rm g})  I^{\per,\text{reg}}_{1/(r - r_i)}(r_{2\rm g})\frac{\delta r_i}{2} + T_{\rm g}(r_{3\rm g})I^\per_{1/(r - r_3)}(r_{2 \rm g})\frac{\delta r_3}{2} + \frac{\delta E}{1 - E^2_{\rm g}} \Big( I^\per_{r^2}(r_{2\rm g}) + 2 I^\per_{r}(r_{2\rm g}) \Big) \nonumber \\
        & + \frac{4\delta E}{1 - E^2_{\rm g}} I^\per(r_{2\rm g}) + \frac{E_{\rm g}}{2}\Big( \displaystyle \sum^3_{i=1} \delta r_i + 2 \delta r_4 \Big) I^\per_r(r_{2 \rm g}) + \frac{E_{\rm g}}{2} \Big(\displaystyle \sum^3_{i=1} (2 + r_{i \rm g})\delta r_i - a + 4\delta r_4 \Big) I^\per(r_{2 \rm g}) \bigg]    \ , \label{eq:shift_coordinate_time_frequency} \\
       \delta \Upsilon_\phi &= \frac{\Upsilon_{\phi \rm g}}{\Upsilon_{r \rm g}} \delta \Upsilon_r + \frac{\Upsilon_{r \rm g}}{\pi}\bigg[ \displaystyle \sum^2_{i=1} \Phi_{\rm g}(r_{i \rm g}) I^{\per,\text{reg}}_{1/(r - r_i)}(r_{2\rm g})\frac{\delta r_i}{2} + \Phi_{\rm g}(r_{3\rm g})I^\per_{1/(r - r_3)}(r_{2\rm g})\frac{\delta r_3}{2} - \bigg(E_{\rm g} - \frac{L_{z \rm g} \delta E}{1 - E^2_{\rm g}} - \delta L_z \bigg) I^\per(r_{2 \rm g}) \bigg]   \ , \label{eq:shift_azimuthal_frequency}
    \end{align}    
\end{widetext}
These expressions are in excellent numerical agreement with the result of Refs.~\cite{Drummond:2022_near_eq,Piovano:2024yks}. Moreover, Eqs.~\eqref{eq:shift_radial_frequency}-\eqref{eq:shift_azimuthal_frequency}  in the FC and FE gauges perfectly match the numerical results for the frequencies obtained with the code of~\cite{Piovano:2024yks,repoHJ}.

\subsection{Frequencies shifts for homoclinic trajectories}
The corrections to the frequencies are finite in the limit $r_{3\rm g} \to r_{2 \rm g}$ only in the fixed eccentricity gauge. In particular, the shift to the radial frequency $\delta \Upsilon^\FE_r$ vanishes, while
\begin{align}
   \frac{\delta \Upsilon^\FE_r}{\Upsilon_{r \rm g}} &= \frac{a}{2r^2_{2\rm g}} - \frac{E_{\rm g} \delta E^\FE }{1 - E^2_{\rm g}} - \frac{\delta r_4}{r_{2 \rm g}}   \nonumber \\
   &+ \frac{r_{2 \rm g} \delta r^\FE_1 + r_{1 \rm g} \delta r^\FE_2 - 2 r_{2 \rm g}\delta r^\FE_2}{2(r_{1 \rm g} - r_{2 \rm g})r_{2 \rm g}}   
\end{align}
and
\begin{widetext}
    \begin{align}
        \delta \Upsilon^\FE_t &= T_{\rm g}(r_{2 \rm g})\frac{\delta \Upsilon^\FE_r}{\Upsilon_{r \rm g}} - \frac{T_{\rm g}(r_{1 \rm g})}{2(r_{1 \rm g} - r_{2 \rm g})}\delta r^\FE_1 - \frac{(r_{1 \rm g} - 2 r_{2\rm g})T_{\rm g}(r_{2 \rm g})}{2(r_{1 \rm g} - r_{2 \rm g})r_{2 \rm g}}\delta r^\FE_2  + \frac{4 + 2r_{2 \rm g} + r_{2 \rm g}}{1 - E^2_{\rm g}}\delta E^\FE  \nonumber \\
        &+ \frac{E_{\rm g}}{2} \big( (2 + r_{1 \rm g} + r_{2 \rm g})\delta r^\FE_1 + 4 (1 + r_{2 \rm g})\delta r^\FE_2 - a + 2(2+r_{2 \rm g})\delta r_4 \big) \ , \\
       \delta \Upsilon^\FE_\phi &= \Phi_{\rm g}(r_{2 \rm g})\frac{\delta \Upsilon^\FE_r}{\Upsilon_{r \rm g}} - \frac{\Phi_{\rm g}(r_{1 \rm g})}{2(r_{1 \rm g} - r_{2 \rm g})}\delta r^\FE_1 - \frac{(r_{1 \rm g} - 2 r_{2\rm g})\Phi_{\rm g}(r_{2 \rm g})}{2(r_{1 \rm g} - r_{2 \rm g})}\delta r^\FE_2 - E_{\rm g} + \frac{L_{z \rm g} \delta E^\FE}{1 - E^2_{\rm g}} + \delta L^\FE_z   \ ,
    \end{align}    
\end{widetext}
All the previous spin corrections to the frequencies are finite in the limit $r_{2 \rm g} \to r_{1 \rm g}$  (i.e. at the geodesic ISCO).

The linear in spin shift to the azimuthal frequency in coordinate time, $\delta \Omega_\phi^\FE$, can be expressed in terms of the Mino-time frequencies in the FE gauge as
\begin{equation}
    \delta \Omega^\FE_\phi = \frac{\delta \Upsilon_\phi^\FE}{\Upsilon_{t \rm g}} - \Upsilon_{\phi \rm g} \frac{\delta \Upsilon_t^\FE}{\Upsilon_{t \rm g}^2}  \ .
\end{equation}
At the geodesic ISCO, $\delta \Omega^\FE_\phi$ perfectly matches the frequency shift found by Jefremov et al in~\cite{Jefremov:2015gza}.


\bibliographystyle{utphys}
\bibliography{Ref}

\end{document}